\documentclass[pdflatex,sn-mathphys-num]{sn-jnl}% Math and Physical Sciences Numbered Reference Style 
\usepackage{graphicx}%
\usepackage{multirow}%
\usepackage{amsmath,amssymb,amsfonts}%
\usepackage{amsthm}%
\usepackage{mathrsfs}%
\usepackage[title]{appendix}%
\usepackage{xcolor}%
\usepackage{textcomp}%
\usepackage{manyfoot}%
\usepackage{booktabs}%
\usepackage{algorithm}%
\usepackage{algorithmicx}%
\usepackage{algpseudocode}%
\usepackage{tabularx, booktabs} % Add this to your preamble
\usepackage{array, booktabs} % Ensure these packages are included in the preamble
\usepackage{listings}%
\usepackage[left]{lineno} % Load lineno package
\usepackage{breqn}
\usepackage{etoolbox}

\theoremstyle{thmstyleone}%
\theoremstyle{thmstyletwo}%

\theoremstyle{thmstylethree}%

\unnumbered% uncomment this for unnumbered level heads

\begin{document}

%\title[Article Title]{Isolator-Free Laser Operation Enabled by Chip-Scale Reflections in Zero-Process-Change SOI}
\title[Article Title]{\makebox[\textwidth][c]{Isolator-Free Laser Operation} \\ \makebox[\textwidth]{Enabled by Chip-Scale Reflections}
\makebox[\textwidth][c]{in Zero-Process-Change SOI}}


\author*[1]{\fnm{Omid} \sur{Esmaeeli}}\email{omdesml@ece.ubc.ca}

\author [1]{\fnm{Lukas} \sur{Chrostowski}}\email{lukasc@ece.ubc.ca}

\author[1]{\fnm{Sudip} \sur{Shekhar}}\email{sudip@ece.ubc.ca}

\affil[1]{\orgdiv{Department of Electrical and Computer Engineering}, \orgname{The University of British Columbia}, \orgaddress{\city{Vancouver}, \state{BC}, \country{Canada}}}


\abstract{The isolation-free operation of photonic integrated circuits enables dense integration, reducing packaging costs and complexity. Most isolator replacements require a change in the silicon-on-insulator (SOI) foundry process and suffer from large insertion loss. Most solutions did not integrate the laser, leaving the verification incomplete, and measurements with modulated reflections have also been missing. In this work, we present, for the first time, a zero-process-change silicon photonic (SiP) circuit that, when paired with an integrated distributed-feedback laser (DFB), enhances the DFB's immunity to continuous-wave and modulated parasitic reflections from multiple reflectors. The circuit generates intentional, controlled self-injection to stabilize laser dynamics and maintain operation. The SiP circuit is complemented by an electro-optic feedback loop that dynamically adjusts the self-injection to preserve laser stability. The proposed circuit introduces an insertion loss of \unboldmath$1.5~\mathrm{dB}$ and enables the DFB laser to tolerate back reflections as large as $-7~\mathrm{dB}$ and $-12~\mathrm{dB}$ from on-chip and off-chip reflectors, respectively. The DFB is hybrid integrated with the SiP chip using a photonic wire bond (PWB). The isolator-free operation of the integrated laser in a high-speed optical link has been demonstrated, highlighting its potential for data communication applications.\newpage}
%%================================%%
%% Sample for structured abstract %%
%%================================%%
\maketitle

\section{Introduction}\label{sec1}
Silicon photonics (SiP) has achieved major industrial success in data centers and is emerging as a leading technology for new applications, leveraging co-integration, advanced packaging, and CMOS manufacturing compatibility \cite{Shekhar2024}. Over the past decade, substantial progress has been made in integrating lasers with silicon through various methods \cite{Han2020b,Caimi2021,Han2020, Jones2019,Jones2021,Zhang2019b,Blaicher2020,Marinins2023,Li2022-qh}. However, industry adoption has relied upon external isolators due to stability concerns of isolator-free operation. 

\begin{figure}[t!]
    \centering        
    \includegraphics[width=\textwidth]{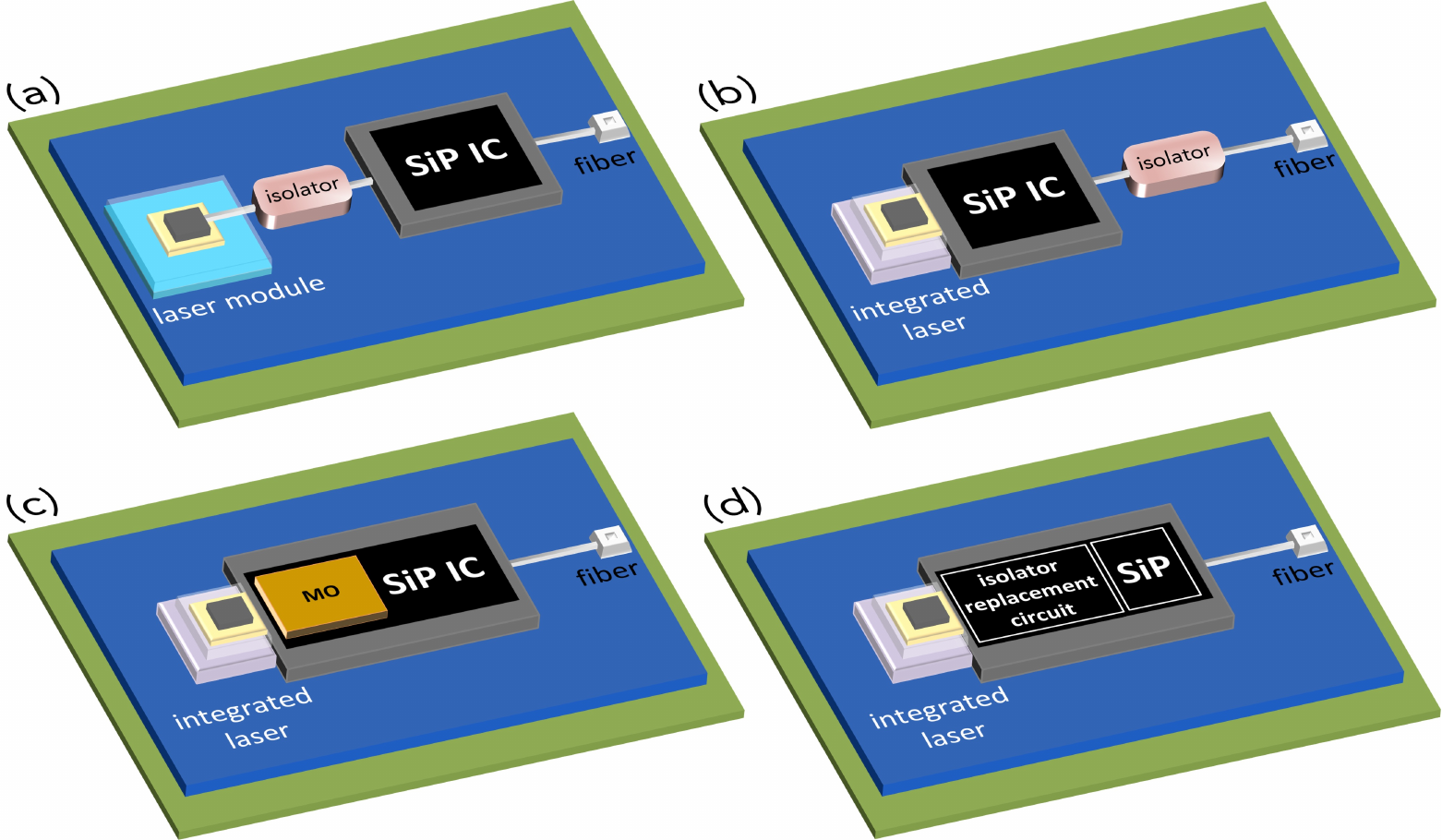}
    \caption{Comparison of different laser-PIC assembly architectures and their isolation strategies. (a) Conventional discrete laser-PIC assembly with an off-chip isolator, ensuring stable operation but increasing size and cost. (b) Integrated laser-PIC assembly with an off-chip isolator positioned at the PIC output, reducing fiber-coupled reflections but leaving the laser sensitive to on-chip reflections and still requiring bulky isolators. (c) Integration of magneto-optic (MO) material on the PIC for on-chip isolation, at the cost of increased fabrication complexity, cost, and insertion loss. (d) Proposed architecture with a silicon photonic (SiP) circuit that enhances laser immunity to reflections, enabling dense integration and cost-effective packaging.}
    \label{fig:arch}
\end{figure}

The laser is exposed to parasitic reflections from the PIC and fiber in a practical photonic integrated circuit (PIC) and laser assembly. Currently, uninterrupted and stable laser operation in advanced PICs is guaranteed by using off-chip isolators (Fig. \ref{fig:arch}(a)). However, these isolators increase the form factor, limit scalability, and significantly raise packaging costs, especially when free-space optics and hermetic sealing are required. Fig. \ref{fig:arch}(b) depicts a commercial architecture where the laser is integrated with the PIC, and the isolator is placed at the PIC output to protect the laser from fiber-coupled parasitic reflections. While fiber-coupled reflections are mitigated, the laser remains susceptible to chip-scale unwanted reflections, requiring careful circuit design to minimize parasitic reflections and often laser design itself. The isolator assembly is still bulky and costly. 
Fig. \ref{fig:arch}(c) shows a diagram where magneto-optic (MO) material has been integrated on the silicon PIC and utilized to break reciprocity and enable on-chip isolation. The increased fabrication cost and complexity have hindered its widespread adoption \cite{Bi2011,Huang2016,Pintus2017,Zhang2019}. Fig. \ref{fig:arch}(d) illustrates the proposed architecture, where the laser is integrated with the PIC, and a SiP circuit is employed to enhance the laser's immunity to reflections. This architecture allows for miniaturized dense electro-optic integration, boosting the performance of the technology in high-value emerging applications.

Employing circuit techniques in native silicon photonics to address the isolation issue is highly desirable due to their simplicity, cost-effectiveness, and compatibility with dense integration. Numerous isolator replacement techniques have been proposed and implemented on the SiP platform.
These solutions can be categorized into two types of zero-process-change, such as employing nonlinear optical effects \cite{Yang2020, Hua2016, Fan2012} or producing spatiotemporal non-reciprocity through index modulation \cite{Lira2012, Kittlaus2021, Kittlaus2018}, and process-change, achieved by integrating magneto-optic materials \cite{Bi2011, Huang2016, Pintus2017,Zhang2019}, or ultra-low-loss SiN process leveraging injection locking \cite{Xiang2023}.

Isolation in nonlinear optical devices relies on the input pump power, with optimal performance limited by the dynamic reciprocity \cite{Shi2015} and the sensitivity of the geometry to fabrication variation. Although this technique offers compact non-reciprocal transmission, its performance remains insufficient in terms of isolation level and insertion loss (IL). IL is a key metric for the practical usefulness of isolator replacements. 

PN-based non-reciprocal devices have been demonstrated on silicon \cite{Pandey21,Tzuang2014,Doerr2014,Doerr2011}. However, this approach requires a large RF drive signal, suffers from high IL, and a large footprint in case of using traveling-wave Mach-Zehnder modulators.

In \cite{Shoman21}, a SiP circuit is introduced that leverages active cancellation of back reflections. Although this design effectively cancels on-chip reflections, it has not been demonstrated for time-evolving fiber-coupled reflections, which are typically observed in real devices. In addition, the loss of the laser-to-PIC coupling was $12~\mathrm{dB}$ (24~dB roundtrip), severely limiting its real-world application.

Self-injection locking (SIL) has been widely studied and primarily pursued as a linewidth reduction technique, leveraging ultra-high Q-factor resonators \cite{Kondratiev2023}. However, its potential as an isolator replacement has been less explored compared to its role in linewidth reduction. Increasing the loaded Q-factor and suppressing the coherence collapse of the laser can make the laser more immune to on-chip and off-chip reflections \cite{Gomez2020,Xiang2023}. In \cite{Xiang2023}, an SIL laser integrated with an ultra-low-loss SiN cavity demonstrated tolerance to $-6.9~\mathrm{dB}$ and $-10~\mathrm{dB}$ of on-chip and fiber-coupled continuous-wave reflections, respectively. Despite its reflection tolerance, SIL is vulnerable to any mechanism that shifts the resonant frequency of the laser or the external cavity relative to each other. Additionally, the locking dynamics is influenced by the phase of the self-injection, which must be carefully tuned and maintained for long-term stability. Ensuring that the lock condition is maintained would support the broader adoption of SIL as an isolator replacement. While the phase tuning has been performed manually in the prior art\thinspace\cite{Xiang2023,Han2024,Su2022,Tang2022}, we have automated the process by incorporating an electro-optic feedback loop. Moreover, since SIL operates by aligning the laser at the resonance of the external cavity, it typically introduces considerable IL.

In this work, we propose a SiP circuit that essentially functions as an adjustable on-chip reflector, providing control over laser dynamics. The proposed laser-PIC architecture demonstrates immunity to continuous and time-varying (modulated) parasitic reflections from multiple reflection points while achieving low IL. Additionally, this circuit is equipped with an electro-optic feedback loop that initializes the operation and maintains laser stability by dynamically adjusting the controlled self-injection. The paper is organized as follows: the next section explores the dynamics of the laser under parasitic reflections, emphasizing how the interplay of delay, amplitude, and phase influences the laser's performance, and why strong self-injections from a short on-chip reflector are useful for laser stabilization. This section also introduces the proposed SiP circuit, its operating principles, and the electro-optic feedback loop. Section 3, Results and Demonstration, presents measurement results, demonstrating the laser's improved tolerance to parasitic reflections, and assessing the circuit's performance in a high-speed optical link. Finally, Sections 4 and 5 offer concluding remarks.

\section{Laser dynamics and stabilization techniques}
\subsection{Interplay of delay, phase, and amplitude of reflections}
\begin{figure}[b!]
    \centering
    \includegraphics[width=0.9\linewidth]{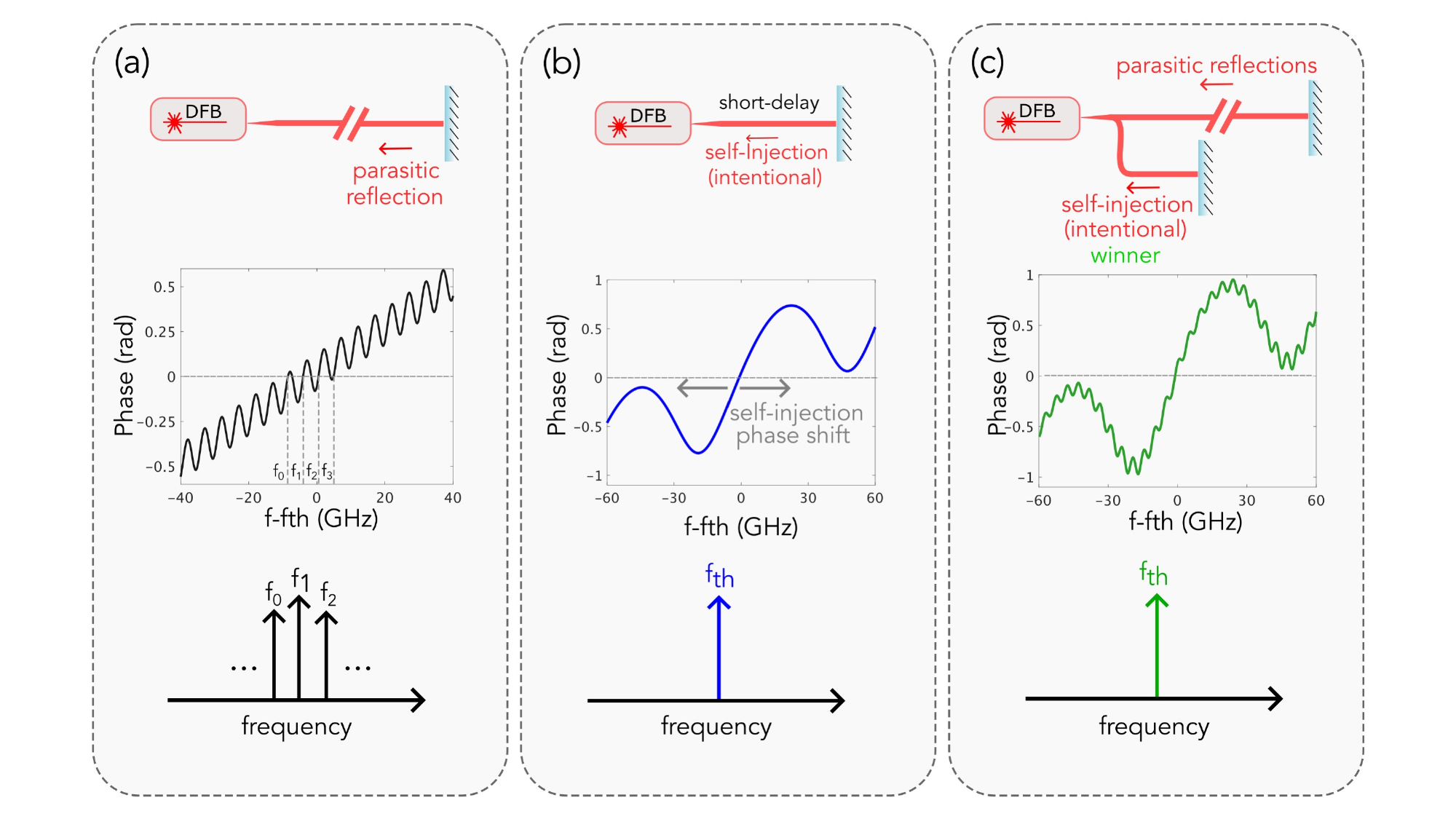}
    \caption{Impact of external cavity reflections on laser stability based on the phase condition ($\Delta\Phi_L$). (a) A distant reflector introduces multiple external cavity modes, destabilizing the laser by causing mode competition. (b) A strong, short-distance reflector increases reflection tolerance, shifting bifurcation points and improving stability. (c) A combination of short and long reflectors, where the strong short reflector dominates laser dynamics, suppressing instability from distant reflections. This highlights the role of controlled self-injection in shaping laser behavior and mitigating parasitic feedback effects.}
    \label{fig:phase}
\end{figure}

In a laser-PIC assembly, numerous reflection points can form, each creating a cavity with the isolator-free laser. The delay of these reflections is the most critical factor in determining the laser's dynamic response. These reflections may originate extremely close to the laser, such as at the coupling edge with PIC, or from distant sources like the output fiber. 
To illustrate the impact of delay and intuitively explain the underlying principle of our approach, the phase of the laser cavity under reflections is plotted in Fig. \ref{fig:phase}. Reflections alter the phase and gain conditions of the laser, potentially resulting in multiple solutions for laser dynamics. For a laser with reflection from an external cavity, the phase condition of the system can be described as (1), where $n_{\mathrm{eff}}$, $L$, and $f$ represent the effective refractive index, length, and emission frequency of the laser cavity, respectively, and c and $m$ denote the speed of light and an integer multiple \cite{Petermann1991}. 
\begin{equation}
    \frac{4\pi n_{\mathrm{eff}} f L}{c} + \phi_{\mathrm{ext}} = 2m\pi
\end{equation}
$\phi_{\mathrm{ext}}$ represents the effective phase of the external cavity. In the absence of external reflections ($\phi_{\mathrm{ext}}=0$), the laser will emit at the threshold frequency ($f_{\mathrm{th}}$). With $\phi_{\mathrm{ext}}=0$, the phase of the laser against frequency follows a straight line that satisfies the phase condition only at the threshold frequency. However, reflections alter the laser frequency, causing the phase condition to deviate from the $2m\pi$ value by $\Delta\Phi_L$. For a laser with multiple external cavities, this can be expressed as \cite{Tager1994,Petermann1991}:
\begin{equation}
    \Delta\Phi_L=2\pi\tau_L(f-f_{\mathrm{th}}) + 2C_l\sqrt{(1+\alpha_H^2)}~\sum_{i} r_{\mathrm{ext,i}} \sin{\left(2\pi\tau_{\mathrm{ext,i}}+\tan^{-1}(\alpha_H)\right)}
\end{equation}
where $\tau_L$, $C_l$, $\alpha_H$, $r_{\mathrm{ext}}$, and $\tau_{\mathrm{ext}}$ represent the laser's cavity roundtrip delay, the output coupling strength of DFB's facet, the linewidth enhancement factor, the external cavity reflectivity, and the external cavity roundtrip time, respectively. 
Fig. \ref{fig:phase} illustrates $\Delta\Phi_L$ of the laser under various external reflectors. The phase condition, along with the gain condition (not shown here), is modulated by the formation of the external cavity. The magnitude and periodicity of the phase deviation depend on the amplitude ($r_{\mathrm{ext}}$) and delay ($\tau_{\mathrm{ext}}$) of the reflection. 

In Fig. \ref{fig:phase}(a), the laser experiences a parasitic reflection from a distant reflector. As a result, multiple external cavity modes satisfy the lasing condition, causing the laser to become unstable. It can be shown that introducing phase shifts to these reflections is insufficient to stabilize the laser \cite{Toomey2015}. In this context, long external cavities cause the most detrimental parasitic reflections. 

In comparison, Fig. \ref{fig:phase}(b) shows the laser's phase condition under a short external reflector, revealing that the laser can tolerate higher levels of reflection before becoming unstable. This behavior has been experimentally demonstrated in the literature, where bifurcation points shift to higher feedback levels in short external cavity regimes \cite{Bosco2017}. The unstable modes also occur at higher frequencies compared to long reflections, and laser stabilization can be achieved by adjusting the phase of the reflections \cite{Schires2017,Esmaeeli2024-qi}. These observations align with our numerical simulation results, provided in Supplementary Section 1, where the stability map of the laser under various reflections has been demonstrated. 

Fig. \ref{fig:phase}(a) and \ref{fig:phase}(b) lead to the proposed concept of using a strong, intentional short reflector as an external cavity to improve the laser's immunity to parasitic reflections. Fig. \ref{fig:phase}(c) shows the phase condition of the laser when subjected to two external cavities. The short external cavity dominates laser dynamics, suppressing the instability caused by the same distant reflections in Fig. \ref{fig:phase}(a).

This approach is made feasible through integrated photonics, enabling reliable and controlled self-injection into the laser. PIC and laser integration eliminates the instabilities in the external cavity that could otherwise occur in free-space optics implementations.

\subsection{Laser-PIC assembly}
\begin{figure}[b!]
    \centering
    \includegraphics[width=1\textwidth,trim= 1.1cm 0 0 0, clip, keepaspectratio]{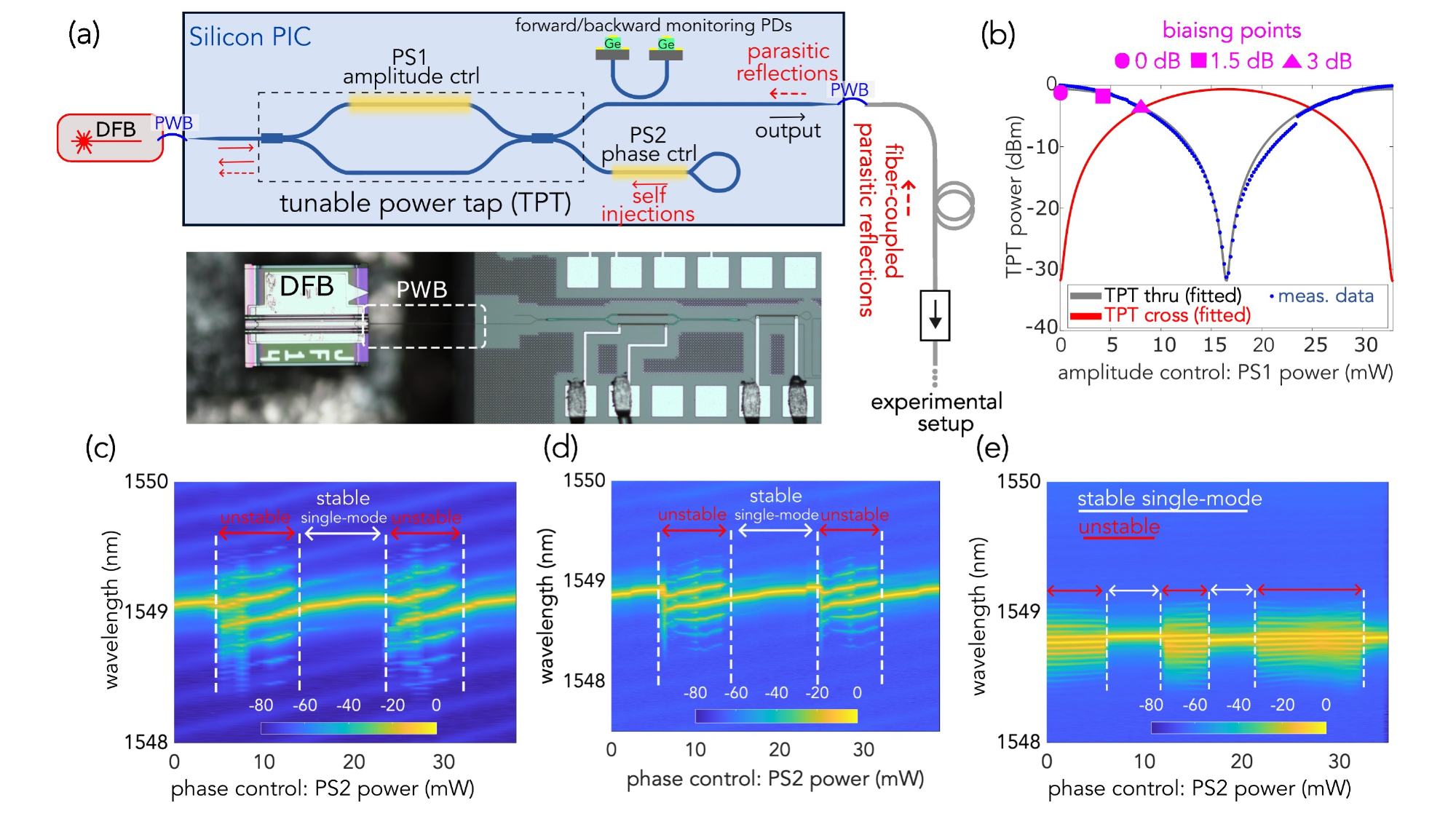}\vspace{-4mm}
    \caption{Characterization of the proposed SiP circuit. (a) (Top) Schematic of the proposed SiP circuit with a loop mirror waveguide and tunable power tap (TPT). PS1 controls the TPT to adjust the self-injection amplitude, and PS2 tunes the phase of the self-injection. (Bottom) Micrograph of the laser-PIC assembly, where the DFB laser is hybrid-integrated with the PIC using a photonic wire bond (PWB). (b) Measured optical power at the output of the PIC as a function of the tunable power tap (TPT), obtained by sweeping the phase shifter PS1. (c-e) Spectrograms of the DFB laser as a function of the phase of self-injection at different TPT biasing points: (c)~3~dB, (d)~1.5~dB, and (e)~0~dB.}
    \label{fig:laser-PIC}
\end{figure}

The schematic of the proposed SiP circuit is shown in Fig. \ref{fig:laser-PIC}(a). The circuit features an intentional on-chip reflector that is implemented as a loop mirror waveguide. A tunable power tap (TPT), configured within an MZI, is positioned between the mirror and the DFB laser. Two thermo-optic phase shifters, PS1 and PS2, enable amplitude and phase control of the optical injections. The resulting on-chip cavity is 930~$\mu$m long, with an approximate roundtrip delay ($\tau_{\mathrm{ext}}$) of 25~ps. The DFB laser is driven with a bias current of 30~mA, outputting an estimated optical power of 3~mW. The laser's relaxation oscillation frequency, $f_{\mathrm{ro}}$, is approximately 8~GHz, yielding a $\tau_{\mathrm{ext}}f_{\mathrm{ro}}$ value of 0.2. For monitoring purposes, photodetectors (PDs) are placed at 5\% power taps in both directions after the TPT.
Fig. \ref{fig:laser-PIC}(b) shows the measured optical output power of the chip as a function of the power consumed by PS1. The TPT can be biased to achieve different levels of self-injection. The level of self-injection is determined by the roundtrip loss of the PWB at the input ($\sim$4~dB) and the TPT bias point.

Fig. \ref{fig:laser-PIC}(c)–\ref{fig:laser-PIC}(e) present the spectrogram of the DFB laser at various TPT biasing points, plotted as a function of the phase of self-injection. In Fig. \ref{fig:laser-PIC}(c) and \ref{fig:laser-PIC}(d), where the TPT biased at 3~dB and 1.5~dB points (producing approximately $-10$~dB and $-15$~dB of self-injections, respectively), the laser operates stably in single-mode or oscillates at external cavity modes, as a result of the short external cavity regime \cite{Schires2017,Tager1994}. The phase of the self-injection acts as a knob to control the laser dynamics and can serve as a turnkey to pulsate the laser. The measurement and calculation of the laser's oscillation frequency at different biasing points of the TPT are provided in Supplementary Section 2. 

As can be seen in Fig. \ref{fig:laser-PIC}(c) and  \ref{fig:laser-PIC}(d), the wavelength of the laser against the phase of reflections changes during the stable period. Since the laser is under strong self-injection, any phase shift in the self-injection path is compensated by the laser's detuning up to the point where the phase shift becomes intolerable by the laser's intracavity oscillations, causing it to oscillate at the external cavity modes.

With weak self-injection, the laser dynamics are no longer dominated by the on-chip cavity, making it vulnerable to parasitics from various sources. In our assembly, the primary source of parasitic reflections is the output PWB located approximately 7~mm from the laser. As shown in Fig. \ref{fig:laser-PIC}(e), the laser spectrum exhibits undamped relaxation oscillations, which resemble a route to chaos dynamics before becoming chaotic. Although self-injection induces intervals of stability, these stable points are highly sensitive and unreliable. Optical spectra and the side mode suppression ratio (SMSR) of the laser in the stabilized region are provided in Supplementary Section 3.

Since a portion of the laser's output power is redirected to generate self-injection, the IL is inherently tied to the level of feedback. To achieve the desired operational regime with minimal IL, we selected the 1.5~dB biasing point as the optimal operating point of the circuit for the remaining experiments, where the laser dynamics are primarily governed by intentional self-injection and the loss is minimized.

\subsection{Electro-optic feedback loop}
Once sufficiently strong self-injection is generated, PS2 can be biased within the stable regions identified in the spectrogram. However, if the relative phase shift between the laser and the on-chip reflector changes due to disturbances, the system may become unstable. To address this, a feedback loop is vital for detecting and mitigating such conditions. In general, stabilization techniques should rely on feedback loops to monitor the laser and restore the locking or stable condition if the system deviates from its optimal operating state.

\begin{figure}[t!]
    \centering
    \includegraphics[width=\linewidth,trim= 0 0 0.5cm 0, clip,keepaspectratio]{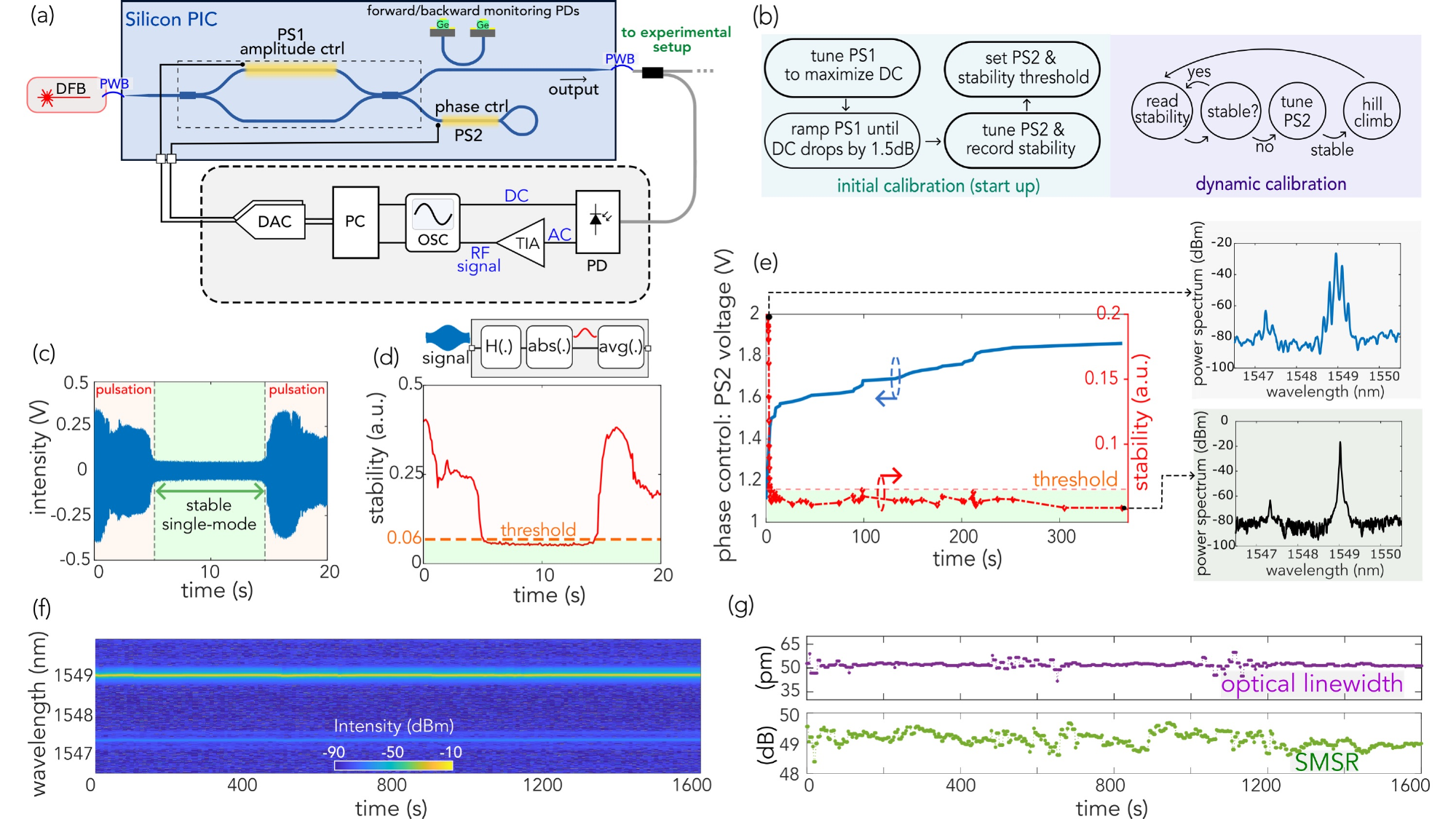}
    \vspace{-2mm}
    \caption{(a) Circuit diagram of the electro-optic feedback loop, where the optical signal from the PIC output is monitored and processed to dynamically adjust self-injection parameters. (b) Feedback algorithm workflow, outlining the initialization and stability optimization processes. (c) Measured RF intensity of the laser at the PIC output as the phase of self-injection is swept, with the TPT biased at 1.5~dB, revealing stable and unstable regions. (d) Extracted stability values derived from RF intensity, used to determine optimal operating conditions. (e) (Left) Measured stability and PS2 drive voltage during real-time feedback loop operation stabilizing the DFB laser. (Right) Optical spectra at the beginning and end of the stabilization. (f) Long-term monitoring of the stabilized laser spectrum. (g) Extracted 20-dB spectral linewidth and SMSR, verifying the effectiveness of the feedback loop in maintaining stable laser operation}
    \vspace{-1mm}
    \label{fig:EO}
\end{figure}

The schematic of the bench-top electro-optic feedback loop is shown in Fig. \ref{fig:EO}(a). The optical signal from the PIC output is tapped for monitoring and routed to an electronic circuit. This circuit includes a photodetector that converts the input light intensity into DC and AC-coupled signals. The AC-coupled signal is processed through a radio-frequency (RF) acquisition path, which includes a transimpedance amplifier (TIA) followed by analog-to-digital converters (ADC) within a real-time oscilloscope. Both the DC and AC signals are then input to the finite state machine (FSM) algorithm implemented on a computer to quantify the stability and execute the feedback mechanism. The algorithm's output is sent to digital-to-analog converters (DAC)% in a source/measure unit (SMU)
, which bias the phase shifters on the PIC. Further details of the electro-optic feedback loop are provided in Supplementary Section 4. 

The feedback algorithm is depicted in Fig. \ref{fig:EO}(b). The initialization starts with finding the 0~dB biasing point (maximum output power) and then ramping up the voltage of PS1 until we bias the MZI at the desired point. Then, the phase shifter is swept, and the stability is recorded. PS2 is biased according to the laser stability.

The stability of the laser is assessed based on its intensity. Fig. \ref{fig:EO}(c) displays the measured AC-coupled intensity of the laser (RF signal) as a voltage ramp is applied to PS2. The unstable regions are characterized by pronounced oscillations at the external cavity modes ($\sim20~$GHz), while the stable single-mode operation is identified by a quiet output intensity dominated by the laser's relative intensity noise (RIN). The oscillation frequency is a function of the delay and strength of the self-injection and is calculated and measured for various cases in Supplementary Section 2.
To quantify stability, the envelope of the recorded RF intensity signal at each step of the phase sweep is extracted using the Hilbert transform, followed by an absolute value operation to generate a positive-valued envelope. The mean of this envelope is then computed using an averaging function. Conceptually, this process is analogous to analyzing the RF spectrum of the intensity signal and calculating the total power of all frequency components.

Fig. \ref{fig:EO}(d) shows the stability values extracted from the signal shown in Fig. \ref{fig:EO}(c). A threshold value, based on the measured stability, is selected to define the boundary of stable operation.
 Fig. \ref{fig:EO}(e) illustrates the feedback loop in action. Initially, PS2 is biased to induce instability in the laser. When the stability value exceeds the threshold, the feedback algorithm activates to tune PS2. During this phase, the voltage actuation is coarse (100~mV) to quickly stabilize the laser, resulting in a sharp initial rise in the voltage (blue curve). Once the stability drops below the threshold (red curve), the algorithm transitions to a hill-climbing optimization (fine tuning) to find the local minimum. In this step, a lower stability threshold is set, and the hill-climbing process continues until the minimum is identified. The optical spectrum of the laser is then recorded for long-term monitoring, as shown in Fig. \ref{fig:EO}(f), with the 20-dB spectral linewidth and SMSR extracted in Fig. \ref{fig:EO}(g). The optical spectra of the laser at the start and end of the feedback operation are also presented. Currently, the feedback loop operates at a speed of approximately 10~Hz, constrained by the computer implementation of the FSM. Employing an application-specific integrated circuit (ASIC) digital design could significantly enhance actuation speed. In addition, ASIC design provides a platform for implementing various feedback algorithms to optimize stability. The algorithm itself is not the primary focus of this work, and it was developed solely to demonstrate the concept of the electro-optic feedback loop.

\section{Results and demonstration}
\begin{figure}[b!]
    \centering
\includegraphics[width=\textwidth, trim= 1cm 0 1.1cm 0, clip, keepaspectratio]{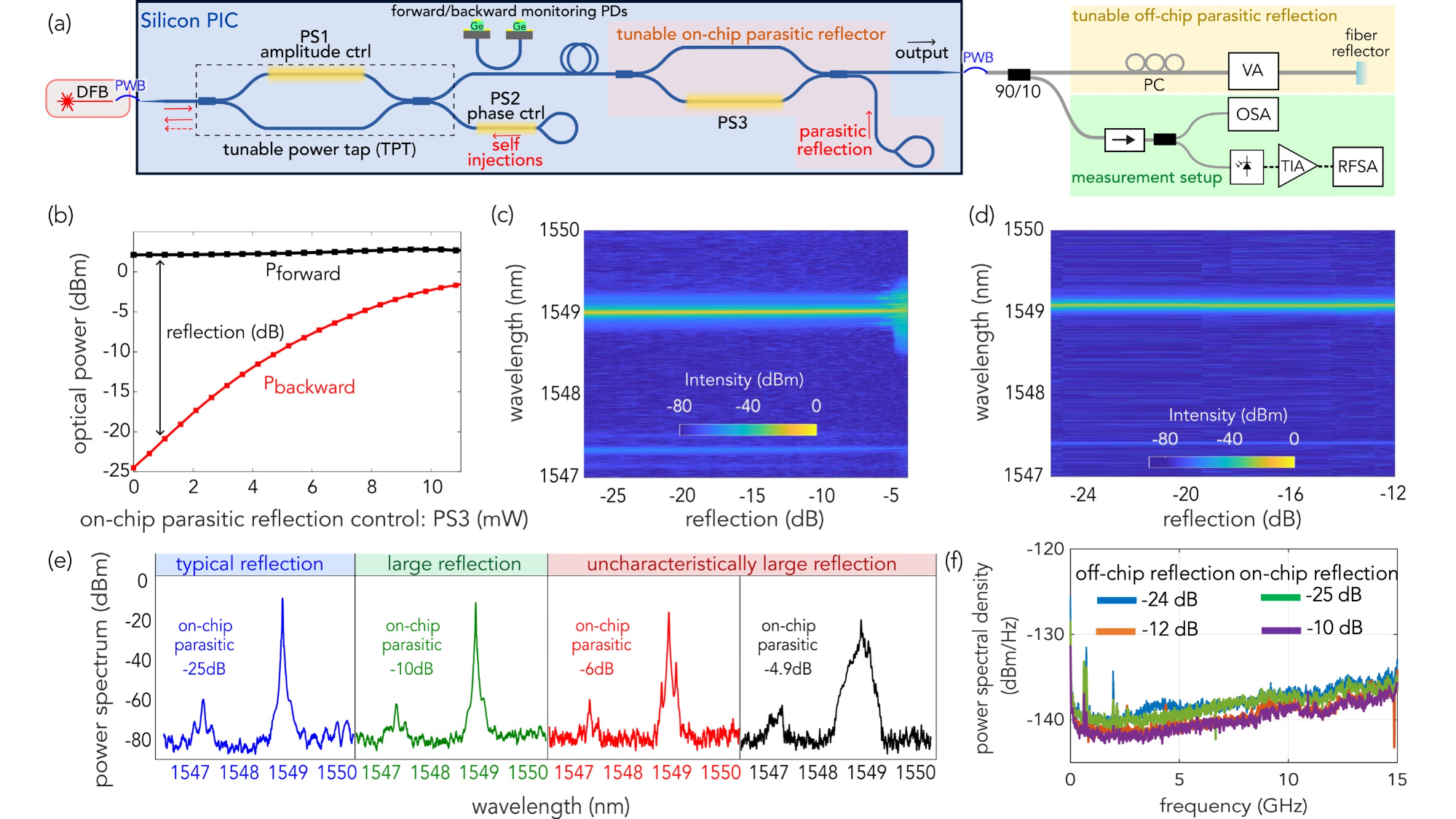}
\vspace{-2mm}
    \caption{(a) Schematic of the SiP circuit and experimental setup used to evaluate the on-chip and off-chip parasitic reflection tolerance of the stabilized DFB laser. (b) Measured forward and backward optical power as a function of the on-chip reflection level, controlled by PS3. (c-d) Spectrograms of the DFB laser under increasing (c) on-chip and (d) off-chip parasitic reflections. (e) Optical spectra of the laser at different on-chip parasitic reflection levels, showing a transition to chaotic operation at (uncharacteristically) higher reflection levels. (f) RF power spectral density of the stabilized DFB laser under the strongest and weakest tolerable parasitic reflection levels from both on-chip and off-chip reflectors.}
    \label{fig:ref}
\end{figure}
\subsection{On-chip and off-chip reflection tolerance}
The stabilized DFB laser with controlled self-injection was tested for tolerance against parasitic on-chip and off-chip (fiber-coupled) continuous wave reflections. Fig. \ref{fig:ref}(a) illustrates the experimental setup. Another on-chip reflector, controlled via an MZI, was incorporated into the circuit to emulate on-chip parasitic reflections. A waveguide delay line was added, making the total distance from the laser to the on-chip reflector 7~mm.

Fig. \ref{fig:ref}(b) shows the measured forward and backward optical power recorded from the on-chip monitoring PDs as a function of PS3, the phase shifter controlling the on-chip parasitic reflection. The level of reflection was measured at the monitoring PDs, located at the output of the self-injection circuit.

The laser was stabilized before generating parasitic reflections, and during the measurements, the amplitude and phase of the self-injection (controlled by PS1 and PS2) were kept constant. Fig. \ref{fig:ref}(c) displays the spectrogram of the DFB laser, measured at the PIC output, as a function of the reflection power. With the controlled self-injection, the laser demonstrated tolerance to back reflections of up to $-7$~dB. The optical spectra of the laser under various reflection levels are shown in Fig. \ref{fig:ref}(e). As the feedback level increases, the external cavity modes become excited (intensity fluctuation in time domain). Further increases in reflection power eventually push the laser into chaotic operation, as indicated by the broadened spectrum observed at approximately $-5~\mathrm{dB}$ reflection.

The tolerance of the stabilized DFB under off-chip, fiber-coupled reflections was evaluated using the test bench shown in Fig. \ref{fig:ref}(a). The parasitic reflector on the chip was bypassed, and the off-chip reflector was created using a $\mathrm{99\%}$ fiber optic reflector in combination with a polarization controller (PC) and a variable attenuator (VA). Fig. \ref{fig:ref}(d) presents the spectrogram of the DFB as a function of the reflection level. The laser remained stable up to $\mathrm{-12~dB}$ of reflection, limited by the roundtrip loss in the setup, with a total fiber length of approximately $10~\mathrm{m}$. The RF power spectral density, illustrating the laser's intensity stability, is shown in Fig. \ref{fig:ref}(f) at both the maximum and minimum levels of reflections. Notably, as shown in Fig. \ref{fig:ref}(c) and (d), the stabilized isolator-free laser does not exhibit a wavelength shift under parasitic reflections, owing to the dominance of the controlled strong self-injection over the intentional external cavity. Testing the feedback tolerance of the free-running laser (i.e., with the TPT bypassed) against the parasitic reflections was impractical, as the laser is unstable due to approximately $-26~\mathrm{dB}$ of reflections generated by the PIC output coupler (edge coupler and PWB interface).

\subsection{Isolator-free operation in high-speed links}
\begin{figure}[b!]
    \centering
    \includegraphics[width=1\linewidth,keepaspectratio]{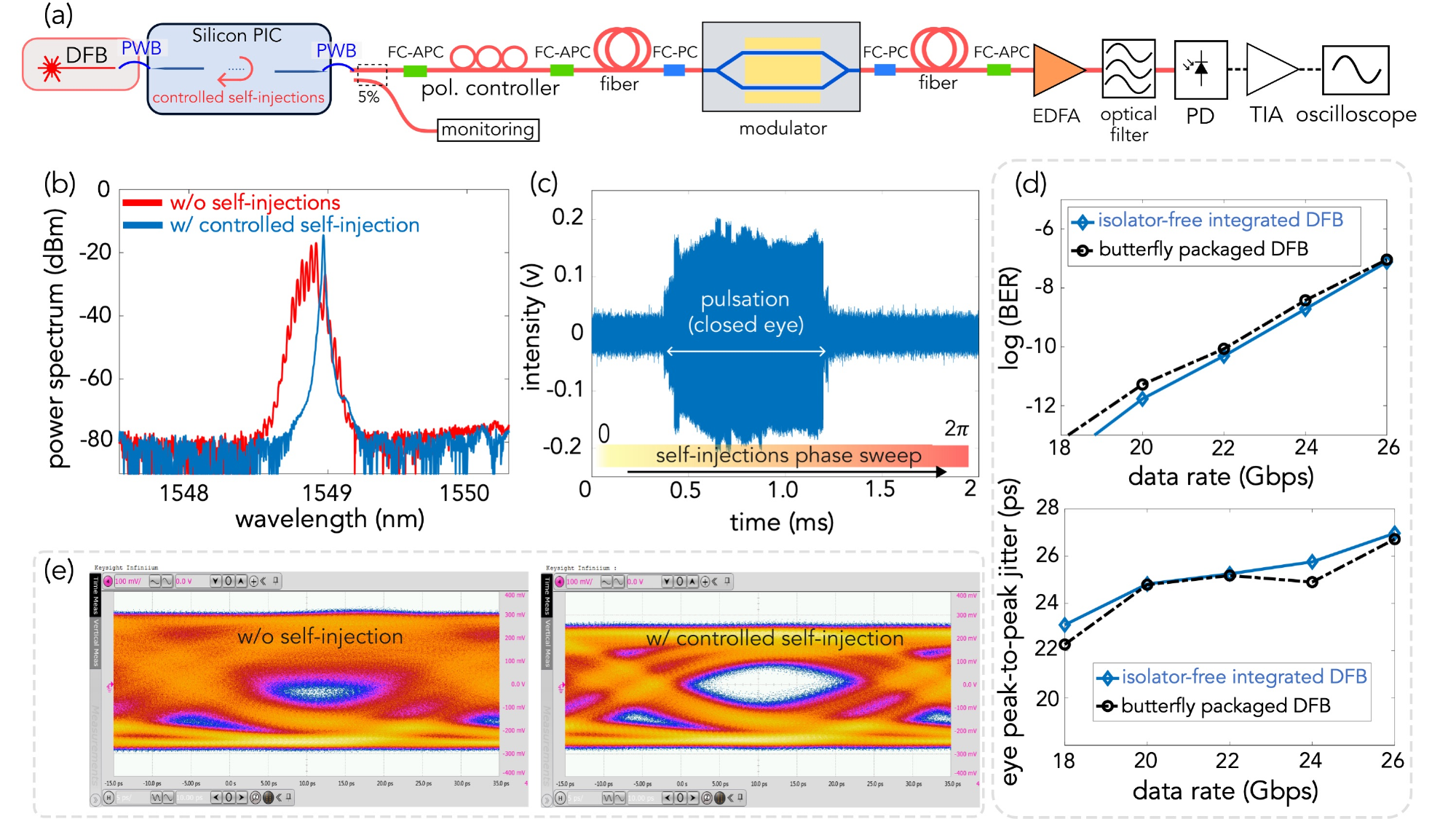}
    \caption{Demonstration of isolator-free laser operation in a high-speed optical link. (a) Isolator-free experimental setup where the integrated laser-PIC assembly is photonic wire bonded to a fiber and connected to an off-chip modulator, with a monitoring tap before the modulator allowing for real-time assessment of the laser.  (b) Measured optical spectrum of the laser before and after stabilization. Without controlled self-injection, parasitic reflections destabilize the laser, but applying the stabilization process achieves stable operation. (c) Measured intensity of the laser during the self-injection phase sweep. (d) Comparison of BER and eye peak-to-peak jitter between the stabilized isolator-free integrated laser and a commercial butterfly-packaged DFB laser with an isolator. (e) Measured PRBS15 25 Gbps eye diagram at the receiver using the integrated DFB laser, with and without controlled self-injection. (FC: Fiber Connector, PC: Physical Contact, APC: Angled Physical Contact.)}
    \label{fig:mod}
\end{figure}
In an optical link, moving from the transmitter (TX) to the receiver (RX), numerous interfaces can generate reflections. These reflections are inherently time-varying due to the modulation applied to the optical signal. In this section, we construct an optical link and evaluate the laser's performance under all parasitic effects that exist in a link. 

In \cite{Matsui2021-ly}, isolator-free operation of a directly modulated quantum-well DFB in a high-speed link has been demonstrated. However, in silicon photonic links, the laser typically operates as a continuous source of light, while the modulation occurs independently. In \cite{Zhang2020}, an isolator-free operation in a coherent link was shown. Nonetheless, the laser was only exposed to continuous-wave back reflections and an explicit isolator was used to prevent modulated back reflections.

 Fig. \ref{fig:mod}(a) depicts the experimental setup for the high-speed optical link. The integrated laser on the SiP chip is photonic wire bonded to a fiber. A $5\%$ monitoring tap at the output of the PIC allows real-time monitoring of the laser's performance before the light is fed to an off-chip modulator. The modulated output is then routed through a fiber coil and delivered to the receiver chain.  

No specific optimizations were made to minimize back reflections in the link, and various types of fiber connectors were used throughout the setup. The laser was exposed to both continuous and modulated back reflections from chip-scale and fiber-coupled sources. To show the effects of parasitic reflections on the laser itself, the TPT was initially bypassed to minimize the self-injection. Fig. \ref{fig:mod}(b) shows the laser spectrum under this condition, revealing instability caused by parasitic reflections. To stabilize the laser, the initial calibration process outlined in Fig. \ref{fig:EO}(b) was applied. During calibration, the laser's intensity was measured while sweeping the self-injection phase, with the TPT biased at 1.5 dB. The results of this process are shown in Fig. \ref{fig:mod}(c). The stabilized laser spectrum is presented in Fig. \ref{fig:mod}(b), confirming the effectiveness of the stabilization. The laser maintained stability throughout the experiments, and the bias voltages for PS1 and PS2 remained unchanged.

The modulator was driven with NRZ PAM-2 signaling at data rates ranging from 18~Gbps to 26~Gbps, limited by the modulator's 10~GHz analog bandwidth. For comparison, the integrated laser and PIC were replaced with a commercial butterfly-packaged DFB laser with an isolator, while keeping the optical power to the polarization controller consistent. The BER results, without any equalization, are shown in Fig. \ref{fig:mod}(d), demonstrating that the isolator-free integrated laser performs similarly to the commercial DFB with an isolator. Peak-to-peak jitter measurements from the eye diagrams are also presented in Fig. \ref{fig:mod}(d). Both measurements show increased BER and jitter at higher data rates due to inter-symbol interference (ISI), as the rates exceed the modulator's 3~dB bandwidth.  

Fig. \ref{fig:mod}(e) compares the 25~Gbps eye diagrams before and after stabilization with self-injection. During link operation, the total back reflection power measured by the on-chip photodetector was $-24$~dBm. The reflections from the off-chip modulator, given its distance, have a more detrimental impact on the laser's stability. Improved performance is anticipated with integrated on-chip silicon modulators. The total fiber length in the setup, excluding instrument delays, was 15~m.

\section{Discussion}

Table I compares the performance of isolator replacement circuits, focusing on silicon-compatible implementations. An SIL technique that leverages ultra-low-loss SiN process \cite{Xiang2023} is also included in the table. The SIL technique and our approach rely on generating desired back reflections to the laser. However, neither our approach nor SIL \cite{Xiang2023} blocks unwanted reflections and, therefore, cannot be characterized by a specific level of isolation. Instead, the maximum tolerable feedback is reported. To enable a comparison, a minimum required isolation can be established. Conventional DFB lasers typically need $<-40~\mathrm{dB}$ of reflection to reliably keep the laser in the single-mode regime of operation. In PIC packaging, the I/O facets, regardless of the edge or surface coupling, can easily generate back reflections in the range of $-25~\mathrm{dB}$ to $-35~\mathrm{dB}$. Downstream on-chip or off-chip components, such as cavities and filters, can produce even higher levels of reflection. Thus, achieving $>30~\mathrm{dB}$ of isolation is necessary to diminish these effects.
\newpage
Quantum dot (QD) lasers warrant a discussion as they present a compelling alternative against quantum-well (QW) lasers. QD lasers have made significant progress as devices, with advancements toward full integration with silicon photonics through both heterogeneous and monolithic approaches \cite{Shang2021-il}. They exhibit a small linewidth enhancement factor and a large damping factor compared to QW-DFBs, making them highly insensitive to optical feedback \cite{Grillot2020-et,Zeyu2019,Huang2018-av}. A recent publication demonstrated the heterogeneous wafer-scale integration of QD-DFBs, achieving a tolerance of up to $-16~\mathrm{dB}$ of back reflection from an on-chip reflector \cite{Duanni2024}. QD lasers have also been demonstrated under continuous-wave reflections in high-speed links \cite{Cui2024-td,Jianan2019,Xian2023}; however, an explicit isolator was still used for the link. Despite all these promising advancements, further isolator-free demonstrations of integrated QD lasers in practical assemblies are needed to fully validate their stability and performance under real-world operating conditions. This includes assessing their maximum tolerance to both chip-scale and fiber-coupled reflections, as well as their viability in isolator-free end-to-end high-speed links. Evaluating the performance enhancement of QD lasers when circuits, such as our proposed approach, are used to generate controlled self-injections would also be of interest.

\begin{table}[t!]
    \caption{Table of comparison for isolator replacements}\label{tab2}
    \renewcommand{\arraystretch}{1.2} % Adjust row height for better spacing
    \setlength{\tabcolsep}{4pt} % Adjust column spacing

    \begin{tabular}{lccccccp{0.5cm}} % Define last column with fixed width
        \toprule
        \begin{tabular}[c]{@{}c@{}}Circuit/\\ Technique\end{tabular} 
        & \begin{tabular}[c]{@{}c@{}}Process\\ Modification\end{tabular} 
        & \begin{tabular}[c]{@{}c@{}}IL\\(dB)\end{tabular} 
        & \begin{tabular}[c]{@{}c@{}}Isolation\\(dB)\end{tabular} 
        & \begin{tabular}[c]{@{}c@{}}Tested w/\\ Integrated\\Laser\end{tabular} 
        & \begin{tabular}[c]{@{}c@{}}Measured \\Reflection\\Tolerance\end{tabular} 
        & \begin{tabular}[c]{@{}c@{}}Tested w/\\ Modulated\\Reflection\end{tabular} 
        & Ref. \\  
        \midrule

        \begin{tabular}[c]{@{}c@{}}Magneto-Optic\\ Ce:YIG/Si\end{tabular} 
        & Yes 
        & \begin{tabular}[c]{@{}c@{}}2.3\\5\\11\end{tabular} 
        & \begin{tabular}[c]{@{}c@{}}32\\30\\32\end{tabular} 
        & No & No & No 
        & \begin{tabular}[c]{@{}c@{}} \cite{Huang2016} \\ \cite{Pintus2017} \\ \cite{Zhang2019} \end{tabular} \\ 

        \addlinespace  

        \begin{tabular}[c]{@{}c@{}}Electro-Optic\\ Modulation\end{tabular} 
        & No  
        & \begin{tabular}[c]{@{}c@{}}11.1\\18\\18\end{tabular} 
        & \begin{tabular}[c]{@{}c@{}}3\\13\\16\end{tabular} 
        & No & No & No 
        & \begin{tabular}[c]{@{}c@{}} \cite{Doerr2014} \\ \cite{Dostart2021} \\ \cite{Pandey21} \end{tabular} \\ 

        \addlinespace  

        \begin{tabular}[c]{@{}c@{}}Optical\\ Nonlinearity\end{tabular} 
        & No  
        & \begin{tabular}[c]{@{}c@{}}1.1--2.5\textsuperscript{1,2}\\15.5\end{tabular}
        & \begin{tabular}[c]{@{}c@{}}20.3--14.2\textsuperscript{1,2}\\40\end{tabular}
        & No & No & No 
        & \begin{tabular}[c]{@{}c@{}} \cite{Yang2020} \\ \cite{Fan2013} \end{tabular} \\  

        \addlinespace  

        \begin{tabular}[c]{@{}c@{}}Reflection\\Cancellation\end{tabular} 
        & No  
        & 3.5 & 16 
        & No & Yes & No 
        & \cite{Shoman21} \\ 

        \addlinespace  

        \begin{tabular}[c]{@{}c@{}}Self-Injection\\Locking\end{tabular} 
        & Yes  
        & 7\footnotemark[1] & NA\footnotemark[3]
        & Yes & Yes & No 
        & \cite{Xiang2023} \\  

        \addlinespace  

        \begin{tabular}[c]{@{}c@{}}Controlled\\ Self-Injection\end{tabular} 
        & No  
        & 1.5 & NA\footnotemark[3]
        & Yes & Yes & Yes 
        & \begin{tabular}[c]{@{}c@{}} this \\ work \end{tabular} \\  

        \bottomrule
    \end{tabular}

    \footnotetext[1]{Extracted from measurement data. \quad  
    \textsuperscript{2}Over input pump power range of 4.9~dBm to 8.5~dBm.}  
    \footnotetext[3]{Not applicable.}  

\end{table}

\section{Conclusion}
In this work, leveraging a PIC implementation, we propose an intentional on-chip short external cavity to stabilize and control laser dynamics. An electro-optic feedback loop is introduced to monitor the laser's stability and adjust the controlled self-injection as needed. The feedback loop quantifies stability based on the envelope of the laser’s intensity and does not require continuous operation. The circuit was tested under various reflection scenarios with a hybrid-integrated DFB laser, demonstrating its ability to maintain stable single-mode operation under parasitic reflections commonly encountered in the practical operation of laser and PIC assemblies. By addressing the isolation challenge with CMOS-compatible solutions, our work enables the compact integration of low-cost DFB lasers, which remain the backbone of the datacom industry, into the next generation of electronic-photonic ICs.

\section*{Methods}
\textbf{Silicon Photonic Circuit:} The SiP chip was fabricated at Advanced Micro Foundry through a standard multi-project wafer (MPW) silicon-on-insulator (SOI) run. The devices are patterned on a 220 nm thick silicon layer. Light is coupled in and out of the optical circuit using inverse nano edge tapers. Thermo-optic titanium nitride (TiN) heaters, with a resistance of approximately 500~$\Omega$, serve as phase shifters. The $2\times2$ couplers in the TPT are rib waveguide adiabatic couplers to minimize parasitic reflections. Monitoring germanium photodetectors with a responsivity of 1.05 A/W are used, operating under a 1~V reverse bias. The output of the PIC is photonic wire bonded to a standard 0-degree angle single-mode fiber array, which has a length of 1~m and is connected to an APC connector.
\vspace{2mm}

\noindent \textbf{Laser-PIC assembly:} The integrated laser is a quantum-well edge-emitting distributed feedback (DFB) laser. Its anti-reflective (AR) coating is index-matched to the printed PWB material, which has an approximate refractive index of 1.5. To characterize the laser, standalone devices from the same batch were photonic wire bonded directly to an optical fiber. Light-current (LI) measurements of these standalone lasers were used to estimate the insertion loss of the laser-to-PIC PWB, which is $\sim2~\mathrm{dB}$. The standalone lasers show a threshold current of approximately 7 mA and a side-mode suppression ratio (SMSR) of 47 dB. The laser and PIC are mounted on a submount to align their heights, with the PIC electrically wire bonded to a breakout PCB. The laser was biased using needle probes, and the submount was temperature-controlled. All measurements were conducted at $25^\circ\text{C}$.
\vspace{2mm}

\noindent \textbf{Electro-optic feedback measurement:} The output of the PIC is connected to a fiber coupler that taps 10\% of the light into the feedback circuit. A Thorlabs 40 GHz RXM40AF photoreceiver is used to convert the optical signal into both DC and AC-coupled (RF) voltage. The photoreceiver is connected to a Keysight DSAX93204A oscilloscope, which captures the data and sends it to a computer running a Python script that implements the FSM algorithm. The PC is also connected to an NI PXIe-6738 DAC, which biases the thermo-optic heaters on the chip.
\vspace{2mm}

\noindent \textbf{On-chip and off-chip reflection tolerance:} The on-chip parasitic reflector is preceded with a waveguide delay line of approximately 5\thinspace mm to make the total distance of the reflector $\sim7$\thinspace mm, mimicking a realistic scenario for silicon photonic circuits. This length results in a $\tau_{\mathrm{ext}}f_{\mathrm{ro}}$ value of approximately 1.4, forming a long external cavity on the chip. In the measurements presented in Fig. \ref{fig:laser-PIC} and Fig. \ref{fig:ref}, an optical isolator was deliberately included to suppress parasitic reflections from the equipment. While these reflections were minimal, the isolator was used to prevent any unintended influence on the laser dynamics. This was important because the measurements were specifically designed to first evaluate the laser dynamics against only the intentional on-chip short reflector in Fig. \ref{fig:laser-PIC}, and subsequently, against both on-chip and off-chip parasitic reflectors in Fig. \ref{fig:ref}. To verify that the isolator had no impact on the observed laser dynamics, the measurements were repeated without it, as detailed in Supplementary Section 3, yielding the same results. An HP~8156A variable attenuator was used for the off-chip reflector, with a minimum transmission loss of 3.1\thinspace dB. The loss of the chip-to-fiber photonic wire bond was measured to be approximately 2.8\thinspace dB. A polarization controller placed before the variable attenuator was used to maximize the backward optical power detected by the on-chip monitoring photodetector. The phase shifters, PS1 and PS2, were biased using Keysight B2901B source/measure unit (SMU) in open-loop mode, without electro-optic feedback constant action. The TPT biasing and adjusting of the self-injection phase were performed automatically. The circuit demonstrated excellent long-term reliability, as the phase shifter bias remained stable throughout all measurements.
\vspace{2mm}

\noindent \textbf{Isolator-free operation in high-speed links:} The off-chip modulator is a Lucent 2623A $\mathrm{LiNbO_3}$ Mach-Zehnder modulator (MZM), biased at the quadrature point using an Anritsu V251 bias-T and Agilent E3631 DC power supply. A PRBS data signal is generated by an Anritsu MU183020A and amplified by an SHF S807C to provide sufficient driving voltage. On the receiving end, an Oprel OFA20D EDFA compensates for losses in the signal path. The amplified optical signal then passes through a Santec OTF900 optical filter before reaching a Thorlabs RXM40AF photoreceiver. Finally, the electrical signal is fed into a Keysight DSAX93204A oscilloscope for data transmission measurements.
\section*{Acknowledgement}
We thank Matthew Mitchell (Dream Photonics) and Shangxuan Yu (UBC) for packaging, Duanni Huang and Haisheng Rong (Intel Corporation) for helpful discussions, CMC Microsystems for fabrication access, and Schmidt Sciences, NSERC and Intel for financial support.

\section*{Data availability}
The supporting data for the conclusions of this investigation are available upon reasonable request.
\section*{Code availability}
The algorithms used for this study are standard and the corresponding authors can provide code scripts upon reasonable request.

\bibliography{sn-bibliography}% common bib file
\pagebreak

\begin{center}
\textbf{\large Supplementary Information: Isolator-Free Laser Operation Enabled by Chip-Scale Reflections in Zero-Process-Change SOI}
\end{center}
\bigskip

\renewcommand{\theequation}{S\thesection.\arabic{equation}}
\renewcommand{\thetable}{S\arabic{equation}}
\setcounter{section}{1} 
\renewcommand{\thefigure}{S\thesection.\arabic{figure}}
\setcounter{figure}{0}

% Patch \cite to add "S" before numbers
\makeatletter
\patchcmd{\@citex}
  {\@citea\hyper@natlinkstart{#2}{\@citeb}}
  {\@citea\hyper@natlinkstart{#2}{S\@citeb}}{}{}
\patchcmd{\@biblabel}
  {#1}
  {S#1}{}{}
\makeatother

\subsection{Section 1: Rate equation based simulation and modeling}
The purpose of this section is to demonstrate through numerical rate equations that the proposed circuit can enhance the stability of the laser against parasitic reflections from short and long external cavities. Our aim is not to replicate the experimental results from the paper, as that would require precise modeling and parameter extraction of the laser. Instead, we will develop a model focused on high-power reflections from multiple external cavity mirrors and simulate the laser using parameters drawn from existing literature. 

As a proof of concept, we intend to identify laser dynamics that align with the behavior observed experimentally through using the Lang-Kobayashi (LK) rate equations that have been used widely to predict laser dynamics under delayed reflections since 1976 [S1]. The equations are originally based upon the small-signal assumption of the reflected power compared to the total emitted power of the laser. The LK rate equations have evolved over the past decades [S2-S7]. We adopt a model proposed in [S2] to capture multiple roundtrips of reflection, which occur due to strong reflections. However, since this model only captures one reflector, we developed our model to simulate our proposed circuit where the laser is under a strong intentional short reflector and parasitic reflections.
\begin{figure}[h!]
    \centering
    \includegraphics[width=0.75\linewidth]{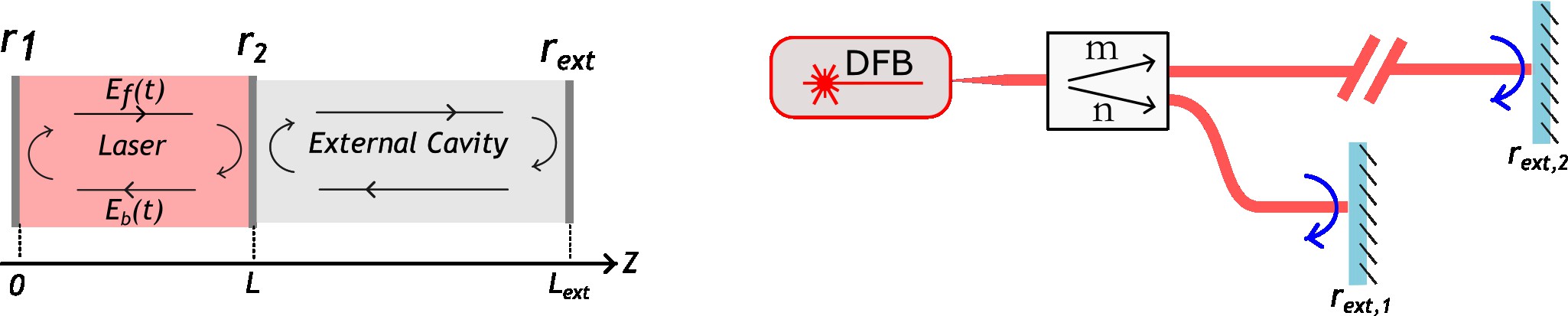}
    \caption{Schematic of the laser under reflections from (Left) a single and (Right) two reflectors.}
    \label{fig:S1}
\end{figure}
The most straightforward approach to model the external reflection is to embed the external cavity(s) transfer function into the laser’s front facet. Fig. \ref{fig:S1} (Left) shows a block diagram of a DFB laser under optical reflection. The laser, with front and back facet reflectivities of $r_1$ and $r_2$, is coupled to a passive external cavity that reflects the emitted output light back into the laser cavity with a reflectivity of $r_{\mathrm{ext}}$. As shown in Fig. \ref{fig:S1}, at the front facet of the laser, based on the stationary ingoing and outgoing complex E-field ($E(t)$), one can write \newline
\begin{equation}
   E_b (t)= r_2 E_f (t) + t_2^2 r_{ext} \sum_{p=1}^{\infty}(-r_2r_{ext})^{p-1}E_f (t-p\tau_{ext}).
   \label{eqn:EbSeries}
\end{equation}
where the $E_f(t)$ and $E_b(t)$ are forward and backward E-fields, with $\tau_{ext}$, $t_2$, and $p$ representing the external cavity roundtrip delay, the laser's front facet E-field transmission coefficient, and the $p^{\mathrm{th}}$ roundtrip, respectively. The effective reflectivity of the front facet ($r_{\mathrm{eff}}=E_b (t)/E_f (t)$) is given by
\begin{equation}
    r_{\mathrm{eff}}=r_2+r_{ext}(1-r_2^2)\sum_{p=1}^{\infty}(-r_2r_{ext})^{p-1}\frac{E_f (t-p\tau_{ext})}{E_b (t)}
    \label{eqn:r2Series}
\end{equation}
Assuming $E(t)=\sqrt{S(t)}e^{-j\{\omega t+\phi(t)\}}$, where \( S(t) \) is the number of photons, \( \omega \) is the emission angular frequency, and \( \phi(t) \) is the phase of the photons, we can substitute this into Eq. (\ref{eqn:r2Series}) and follow the routine in [S7] to calculate \( r_{\mathrm{eff}} \) as
\begin{equation}
    r_{eff}=r_2\Big(\frac{1+\frac{r_{ext}}{r_2}\sqrt{\frac{S(t-\tau_{ext})}{S (t)}}e^{-j\{\omega\tau_{ext}+\phi(t)-\phi(t-\tau_{ext})\}}}{1+r_{ext}r_2\sqrt{\frac{S(t-\tau_{ext})}{S (t)}}e^{-j\{\omega\tau_{ext}+\phi(t)-\phi(t-\tau_{ext})\}}} \Big)=r_2T,
    \label{eqn:reffT}
\end{equation}
where the term multiplied by $r_2$ is the external cavity transfer function, $T$. This coefficient is a complex number and can be represented as $T=|T|e^{-j\angle(T)}$.

The multiple external cavity reflection model is developed in the literature [S4]. However, the multiple roundtrip effect inside the external cavities is neglected. Here, we develop the model for high-power feedback simulations. Furthermore, we take into account the leakage of reflections between mirrors for precision. Fig. \ref{fig:S1} (Right) shows the schematic of the proposed model.

Since all the reflectors share one longitudinal path toward the laser cavity, the emitted power has to be split between reflectors. This has been captured through $m$ and $n$ E-field coefficients with the assumption of $m^2+n^2=1$. Similar to Eq. (\ref{eqn:EbSeries}), by writing E-field equations at the front facet and applying the same steps used in Eq. (\ref{eqn:EbSeries})-(\ref{eqn:reffT}), the transfer function coefficient of the effective reflectivity can be expressed as
\begin{dmath}
        T=1+\frac{(1-r_2^2)}{r_2}\Big[\frac{m^2r_{ext,1}\sqrt{\frac{S(t-\tau_{ext,1})}{S (t)}}e^{-j\{\omega\tau_{ext,1}+\phi(t)-\phi(t-\tau_{ext,1})\}}}{1+m^2r_{ext,1}r_2\sqrt{\frac{S(t-\tau_{ext,1})}{S (t)}}e^{-j\{\omega\tau_{ext,1}+\phi(t)-\phi(t-\tau_{ext,1})\}}}+\frac{n^2r_{ext,2}\sqrt{\frac{S(t-\tau_{ext,2})}{S (t)}}e^{-j\{\omega\tau_{ext,2}+\phi(t)-\phi(t-\tau_{ext,2})\}}}{1+n^2r_{ext,2}r_2\sqrt{\frac{S(t-\tau_{ext,2})}{S (t)}}e^{-j\{\omega\tau_{ext,2}+\phi(t)-\phi(t-\tau_{ext,2})\}}}+2\{\frac{m^2r_{ext,1}\sqrt{\frac{S(t-\tau_{ext,1})}{S (t)}}e^{-j\{\omega\tau_{ext,1}+\phi(t)-\phi(t-\tau_{ext,1})\}}}{1+m^2r_{ext,1}r_2\sqrt{\frac{S(t-\tau_{ext,1})}{S (t)}}e^{-j\{\omega\tau_{ext,1}+\phi(t)-\phi(t-\tau_{ext,1})\}}}\}\times\{\frac{n^2r_{ext,2}\sqrt{\frac{S(t-\tau_{ext,2})}{S (t)}}e^{-j\{\omega\tau_{ext,2}+\phi(t)-\phi(t-\tau_{ext,2})\}}}{1+n^2r_{ext,2}r_2\sqrt{\frac{S(t-\tau_{ext,2})}{S (t)}}e^{-j\{\omega\tau_{ext,2}+\phi(t)-\phi(t-\tau_{ext,2})\}}}\}\Big],
\label{eqn:multi}        
\end{dmath}
where the first two terms in the square brackets account for each mirror's reflectivity ($r_{\mathrm{ext,1}}$ and $r_{\mathrm{ext,1}}$), and the third term captures the leakage of reflections from one reflection path to the other due to the laser facet reflectivity. By plugging this transfer function coefficient, $T$, into the rate equations, the laser under multiple high-power reflectors can be simulated.

The rate equations of the number of photons ($S(t)$), number of carriers ($N(t)$), and phase of photons ($\phi(t)$) can be written based on the modified effective front facet reflectivity ($r_{\mathrm{eff}}$) [S2].

\begin{equation}
    \frac{dS(t)}{dt}=\big\{\frac{G(N,N_{tr})}{1+\epsilon S(t)}-\frac{1}{\tau_{ph}}+\frac{2ln(T)}{\tau_D} \big\}S+\frac{\beta N}{\tau_n}+F_S(t)
    \label{eqn:rateS}
\end{equation}
\begin{equation}
    \frac{dN}{dt}=\frac{\eta I}{q}-\frac{N}{\tau_n}-\frac{G(N,N_{tr})}{1+\epsilon S(t)}S(t)+F_N(t)
    \label{eqn:rateN}
\end{equation}
\begin{equation}
    \frac{d\phi(t)}{dt}=\frac{1}{2}\alpha_H G(N,N_{th})+\frac{arg(T)}{\tau_D}+F_{\phi}(t)
    \label{eqn:ratePh}
\end{equation}

The laser parameters are described in Table \ref{tab:params}. All the laser parameters are taken from [S8], where a DFB laser is numerically simulated, and the laser parameters are calculated for the best experimental and analytical fit outcome.
\renewcommand{\thetable}{S\arabic{table}}
\setcounter{table}{0}
\begin{table}[t!]
\vspace{-2.5mm}
\centering
\caption{Laser parameters description and values}
\begin{tabular}{ll|l}
\hline
\multicolumn{1}{l|}{$G(N,N_{tr})$}      & Active region gain & {$=\partial G/\partial N\times(N-N_{tr})$}              \\ \hline
\multicolumn{1}{l|}{$N_{tr}$}      & Transparency carrier number & {$=8.2\times10^6$}               \\ \hline
\multicolumn{1}{l|}{$N_{th}$}      & Number of carriers above threshold & {$=7.066\times10^7$}              \\ \hline
\multicolumn{1}{l|}{$\partial G/\partial N$} & Active region gain slope constant & {$=1.13\times10^4~(s^{-1})$}                     \\ \hline
\multicolumn{1}{l|}{$\epsilon$}  & Gain compression coefficient & {$=4.58\times10^{-8}$}             \\ \hline
\multicolumn{1}{l|}{$\tau_{ph}$}       & Photon lifetime & {$=7.15~(ps)$}                          \\ \hline
\multicolumn{1}{l|}{$\tau_D$}       & Laser cavity roundtrip delay  & {$=6~(ps)$}            \\ \hline
\multicolumn{1}{l|}{$\beta$}     & Spontaneous emission coupling coefficient & {$=3.55\times10^{-5}$} \\ \hline
\multicolumn{1}{l|}{$\tau_n$}       & Carrier lifetime & {$=0.33~(ns)$}                          \\ \hline
\multicolumn{1}{l|}{$\alpha_H$}     & Linewidth enhancement factor  & {$=4.86$}            \\ \hline
\multicolumn{1}{l|}{$\eta$}      & Quantum efficiency   & {$=0.9$}                     \\ \hline
\multicolumn{1}{l|}{q}        & Electron charge   & {$=1.602\times10^{-19}$(C)}                        \\ \hline
\multicolumn{1}{l|}{$L_D$}       & Laser cavity length & {$=250~(\mu m)$}                      \\ \hline
\multicolumn{1}{l|}{$R_2$}       & Front facet intensity reflectivity & {$=0.3$}       \\ \hline
\multicolumn{1}{l|}{$I$}       & Pump current  & {=26~mA}      \\ \hline
\multicolumn{1}{l|}{$F_S(t),F_N(t),F_{\phi}(t)$}         & Langevin noise terms [S8] & {$NA$}                     \\ \hline
\end{tabular}
\label{tab:params}
\end{table}

\begin{figure}[b!]
    \centering
    \includegraphics[width=0.8\linewidth]{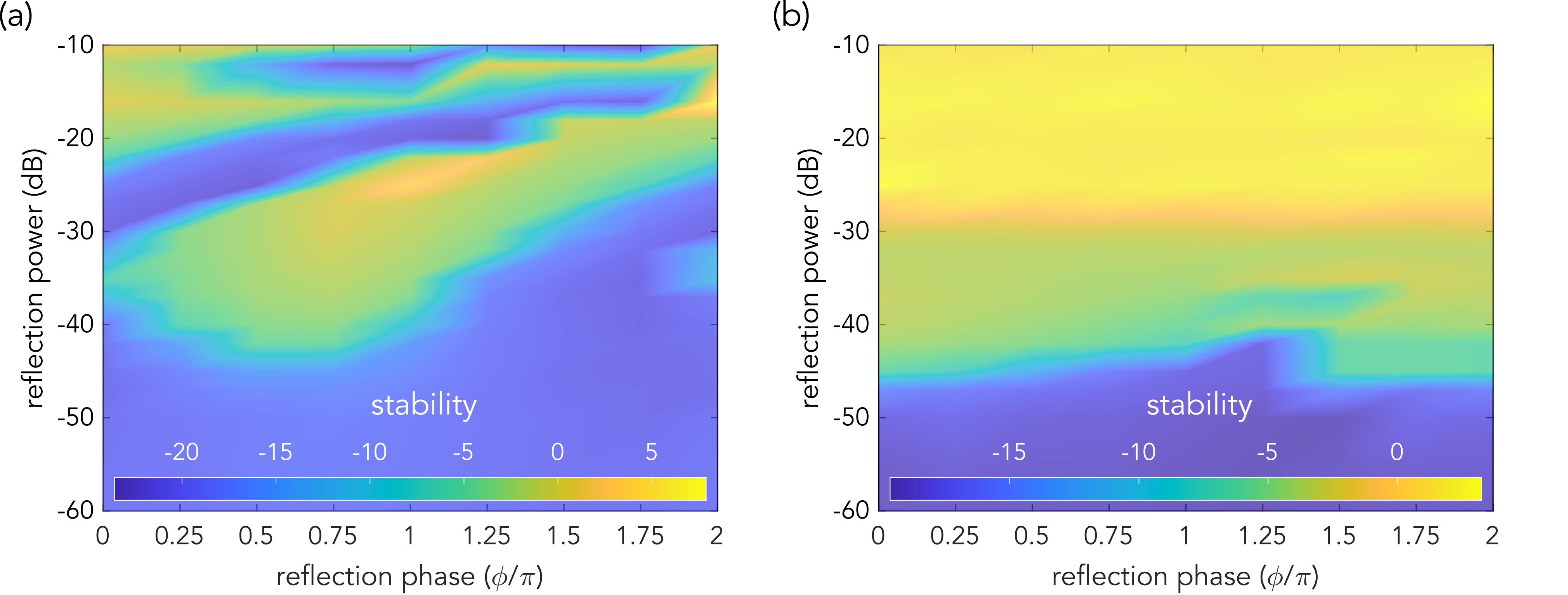}
    \caption{Simulated stability map of the laser under phase and amplitude swept parasitic reflections from a (a) 3~mm and (b) 60~mm reflector.}
    \label{fig:single-ref}
\end{figure}

The stability of the laser against parasitic reflections from both short and long reflectors is numerically simulated and depicted in Fig. \ref{fig:single-ref}. The stability parameter, as described in the main paper, is calculated using the time-domain envelope of the normalized number of photons (S), which represents the intensity of the laser's output. However, to better contrast stable and unstable regions, we take the logarithm of the stability factor. The blue regions indicate stable laser operation, where the photon numbers (S) exhibit minimal fluctuation. The roundtrip delays of the reflectors are 70~ps and 1.4~ns, respectively, resulting in $\tau_{\mathrm{ext}}f_{\mathrm{ro}}$ values of 0.45 and 9. The stability map confirms the phase sensitivity of the laser dynamics to short parasitic reflections. As shown, the laser can be stabilized at high-power reflections by utilizing the phase of the reflections. However, Fig. \ref{fig:single-ref}(b) demonstrates that the laser cannot achieve stability against long reflections through just phase control.

\begin{figure}[h!]
    \centering
    \includegraphics[width=0.8\linewidth]{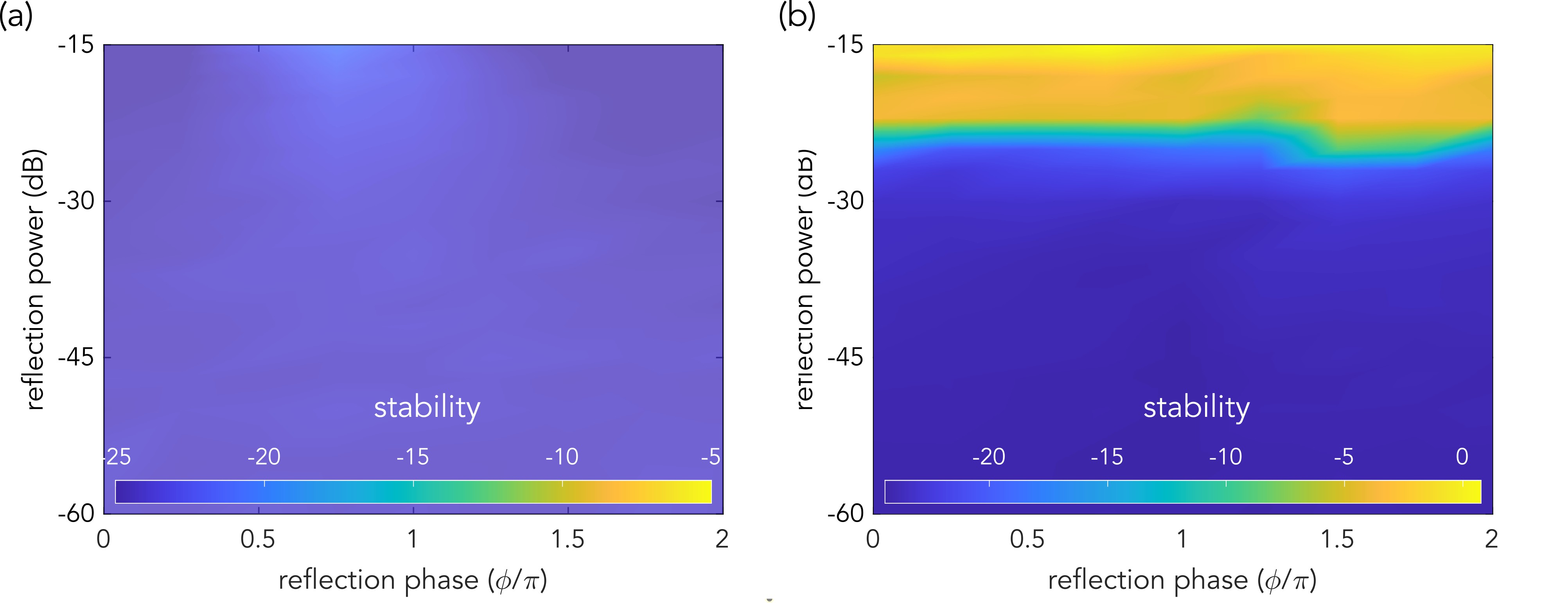}
    \caption{Simulated stability map of the laser stabilized with $-10~\mathrm{dB}$ self-injection from an external short cavity and under phase and amplitude swept parasitic reflections from a (a) 3~mm and (b) 60~mm reflector.}
    \label{fig:SI}
    \vspace{-3mm}
\end{figure}

Using the model from Eq. (\ref{eqn:multi}), we simulated the laser while incorporating the same parasitic reflectors, but added an intentional short external cavity with a length of \( L_{\mathrm{ext}} = 900~\mu m \) that injects reflections at \(-10~\mathrm{dB}\). The stability map of the laser is presented in Fig. \ref{fig:SI}. Comparing these results with those in Fig. \ref{fig:single-ref}, we observe that the laser has become stable against parasitic reflections from the 3 mm parasitic reflector. Additionally, there is an approximate improvement of 30~dB in stability against the long parasitic reflections. 

\setcounter{section}{2} 
\subsection{Section 2: Laser oscillation at external cavity modes}
As shown in the spectrogram of the DFB laser in Fig. 3(c) and 3(d) of the main manuscript, the DFB laser experiences sustained oscillations at the external cavity modes when subjected to the self-injection from a short external cavity with certain phase shifts. This phenomenon arises from the beating of stable and unstable steady-state solutions of the laser with a short external cavity. Here, we utilize equations derived in [S9] to calculate the oscillation frequency of the laser in our system at different biasing points, and compare it with experimental results.
\setcounter{figure}{0}
\begin{figure}[h!]
    \centering
    \includegraphics[width=1\linewidth]{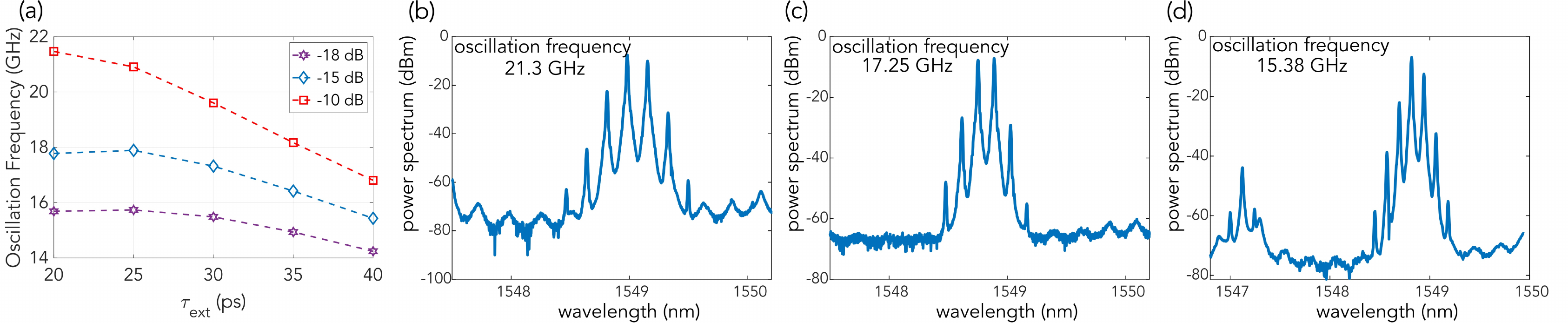}
    \caption{(a) Calculated oscillation frequency of the unstable laser with a short external cavity. Measured optical spectrum of the integrated DFB laser with the intentional self-injection values of (b) $-10~\mathrm{dB}$, (c)$-15~\mathrm{dB}$, and (d) $-18~\mathrm{dB}$.}
    \label{fig:osc}
\end{figure}

Fig. \ref{fig:osc}(a) presents the calculated oscillation frequency values as a function of the roundtrip delay of the external cavity for three different self-injection levels. The length of the intentional on-chip cavity is approximately $930~\mu\mathrm{m}$, resulting in a $\tau_{\mathrm{ext}}$ value of approximately 25~ps. The reflection values of $-10$~dB and $-15$~dB correspond to biasing the TPT at the 3~dB and 1.5~dB points, respectively. Figs. \ref{fig:osc}(b)-(d) illustrate the optical spectrum of the laser oscillating at the external cavity modes at various TPT biasing points. The frequency of oscillation is determined by the wavelength difference between the observed peaks in the optical spectrum, and these measurements align closely with the calculated values. 
\par
\begin{figure}[h!]
    \centering
    \includegraphics[width=0.7\linewidth]{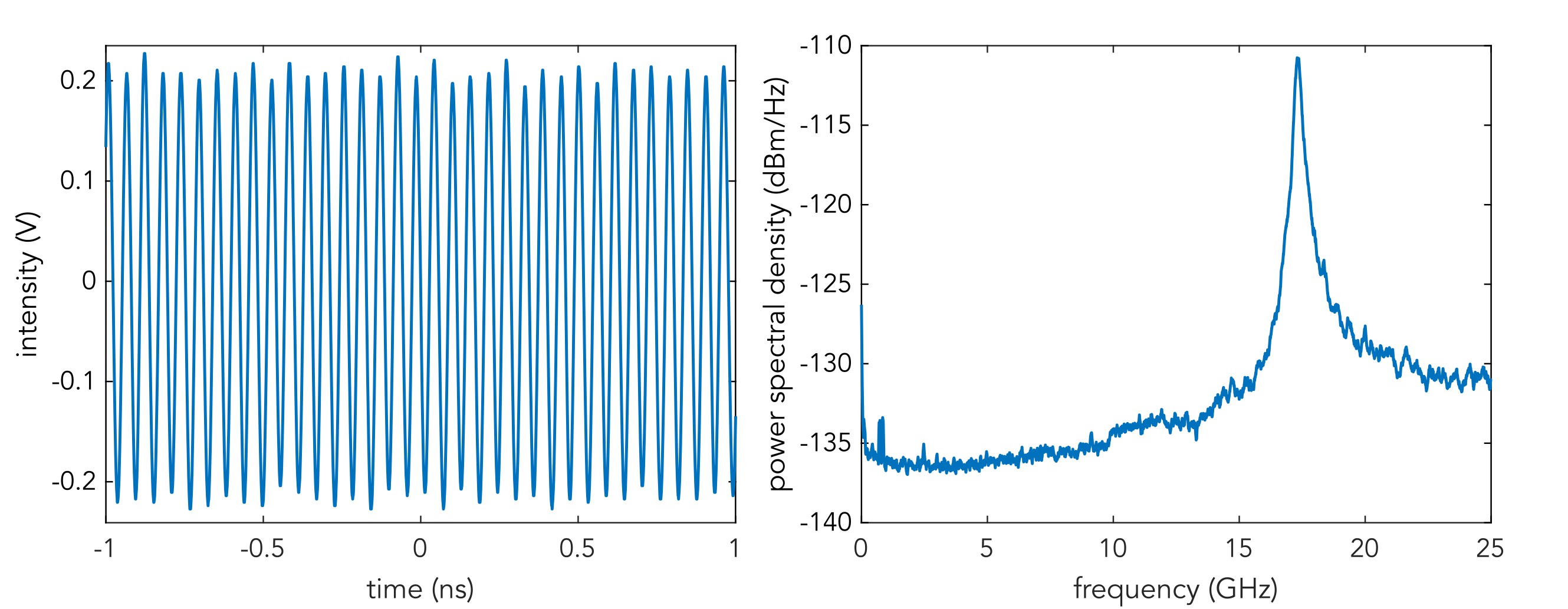}
    \caption{Measured (Left) time-domain intensity and (Right) RF spectral density of the integrated laser at the output of the PIC while it oscillates with the TPT biased at 1.5~dB.}
    \label{fig:osc-1p5dB}
\end{figure}

Fig. \ref{fig:osc-1p5dB} presents time-domain intensity and RF spectral density measurements of the integrated laser at the PIC output when the DFB oscillates at the external cavity modes, emitting the optical spectrum shown in Fig. \ref{fig:osc}(c).

\setcounter{section}{3}
\subsection{Section 3: Optical spectra of the stabilized laser}
Fig. \ref{fig:spectrogram} depicts the measured spectrogram of the DFB laser, showcasing the laser's dynamics as a function of the phase of self-injection at the TPT bias of 1.5~dB. This setup reproduces that of Fig. 3 in the main manuscript without any isolator in the experimental setup. Fig. \ref{fig:spectrogram} depicts a single cycle of the dynamics, occurring over roughly 20 mW of electrical power consumed in PS2.
\setcounter{figure}{0}
\begin{figure}[h!]
    \centering
    \vspace{-5mm}
    \includegraphics[width=0.5\linewidth]{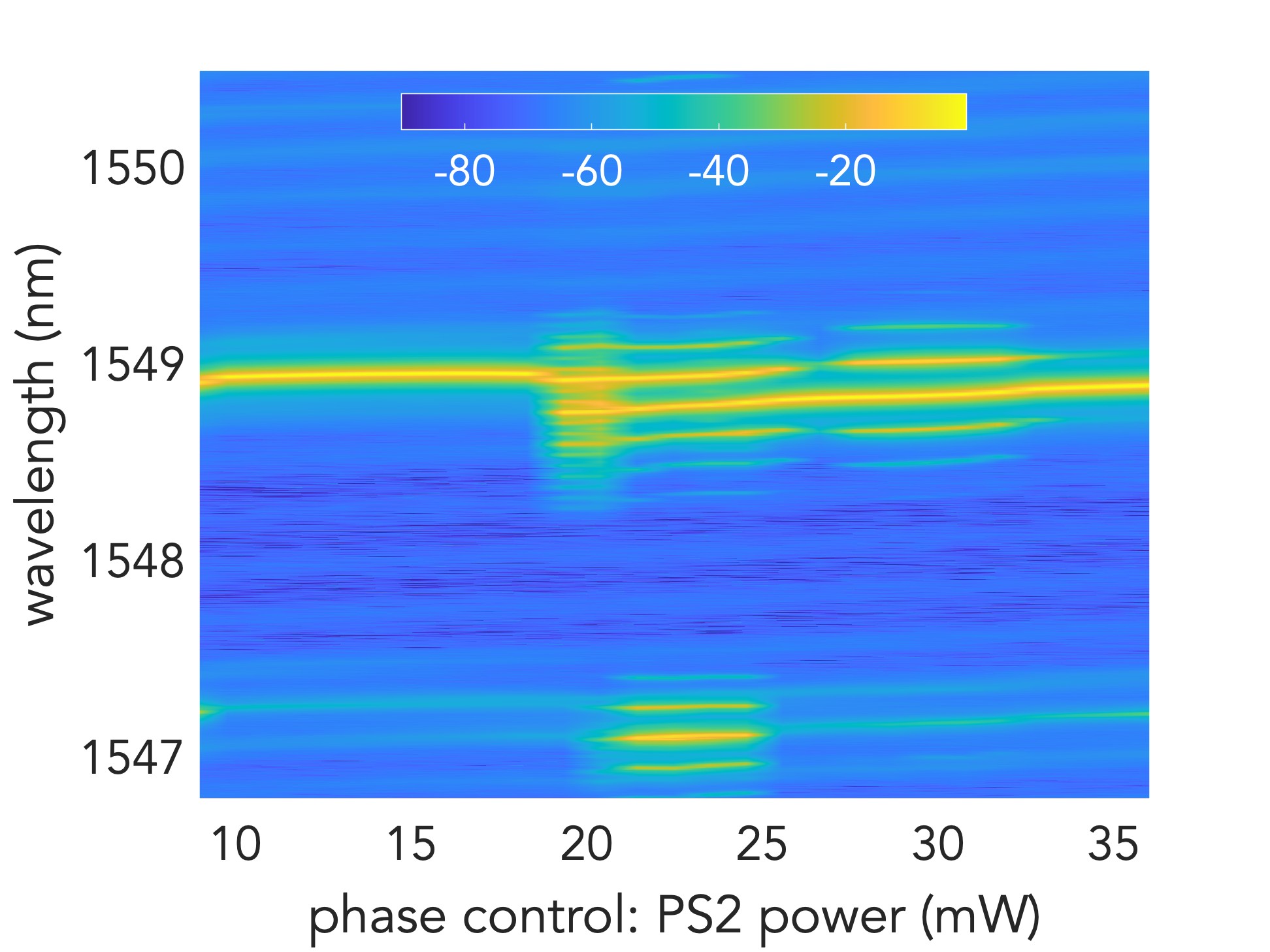}
    \caption{Spectrogram of the DFB laser as a function of the self-injection phase, controlled by PS2, at 1.5~dB biasing point of the TPT.}
\vspace{-3mm}
    \label{fig:spectrogram}
\end{figure}

\begin{figure}[h!]
    \centering
    \vspace{-2mm}
    \includegraphics[width=0.85\linewidth]{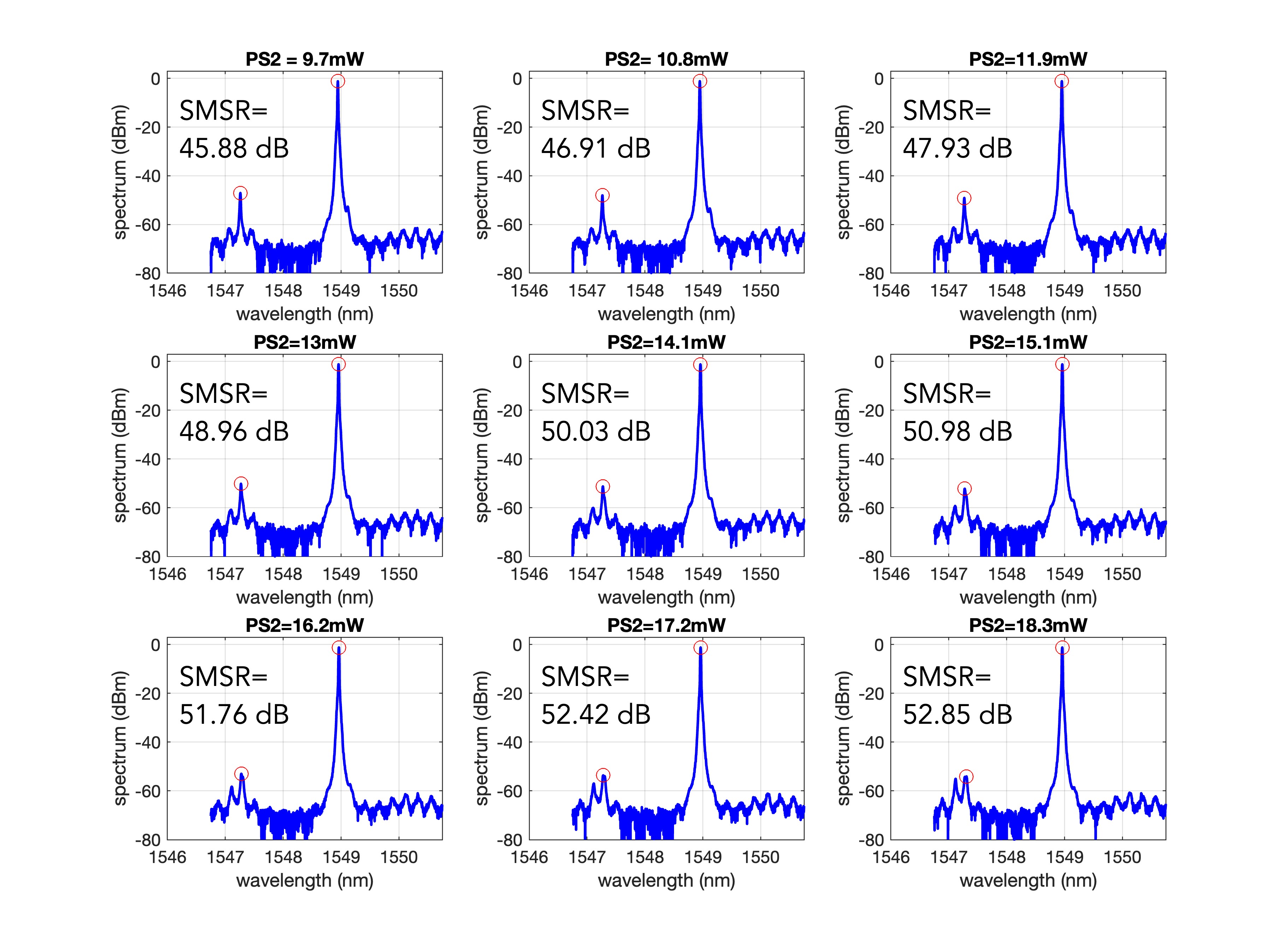}
    \caption{Optical spectra of the stabilized DFB laser under various phase shifts of the self-injection.}
    \label{fig:SMSR}
\end{figure}

Fig. \ref{fig:SMSR} presents the optical spectrum of the DFB laser at various self-injection phase shifts during the stable period. This measurement provides a more detailed view of the optical spectrum and the side mode suppression ratio (SMSR) throughout the stable single-mode operation of the laser. 
\setcounter{section}{4} 
\subsection{Section 4: Electro-optic feedback loop}
The bandwidth of the signal acquisition path (PD, TIA, and oscilloscope), shown in Fig.~4 of the main manuscript, is 33~GHz. The feedback loop can be implemented using commercial off-the-shelf (COTS) components. Figure~\ref{fig:EO-cots} presents a block diagram proposing a board-level electronic feedback circuit. The photodetector, TIA, and analog envelope detector are the only high-speed components in the feedback chain that must support frequencies up to the laser's oscillation frequency. For instance, the Analog Devices ADL602 envelope detector can be used to feed the ADC. The control signal at the output of the envelope detector is low-speed and can be processed using kilohertz-range ADCs before being fed to a microcontroller. A key advantage of the proposed method is that a low-speed control signal is adequate to stabilize the laser. Sub-sampling techniques could be further explored to reduce the high-speed front-end requirements.
\bigskip
\setcounter{figure}{0}
\begin{figure}[h!]
    \centering
    \vspace{-5mm}
    \includegraphics[width=0.6\linewidth]{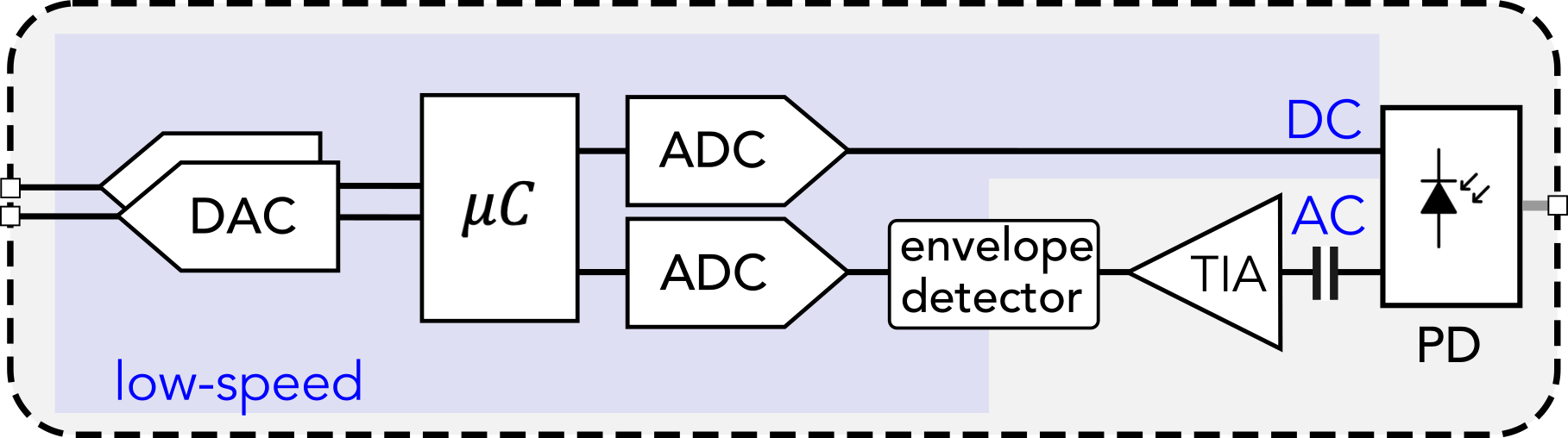}
    \caption{Block diagram of the proposed board-level electronic feedback circuit.}

    \label{fig:EO-cots}
\end{figure}
\vspace{-5mm}

\setcounter{section}{5}
\subsection{Section 5: Linewidth measurement}
\setcounter{figure}{0}
\begin{figure}[h!]
    \centering
    \includegraphics[width=0.6\linewidth]{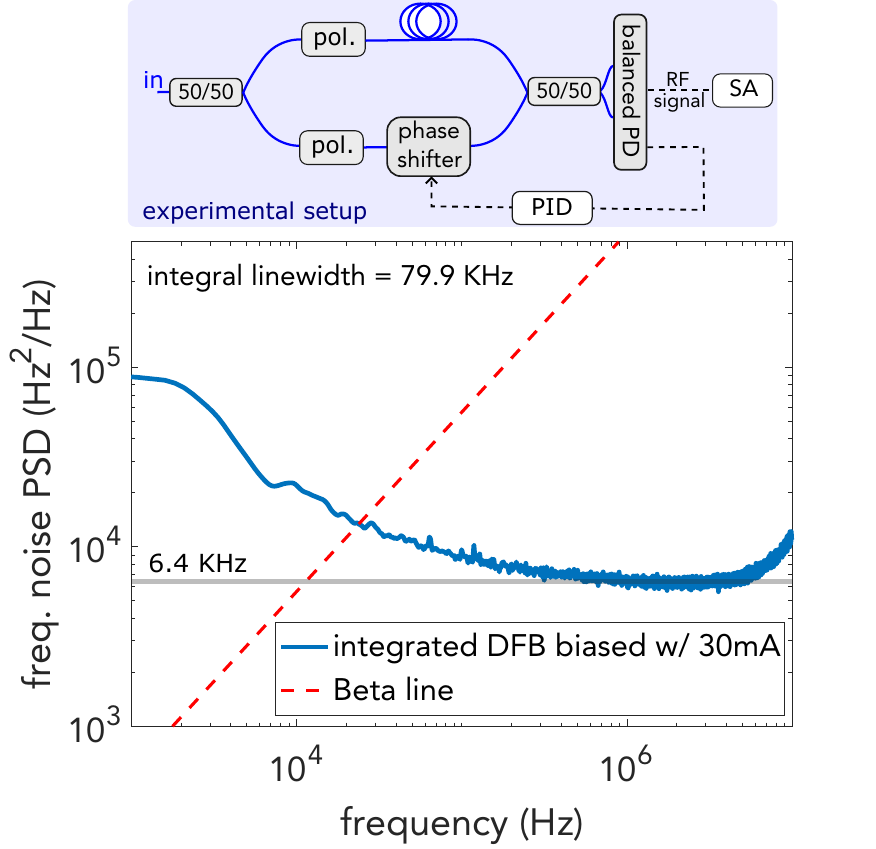}
    \caption{(Top) Experimental setup of the frequency noise discriminator. (Bottom) Linewidth measurement of the stabilized DFB laser.}
    \vspace{-5mm}
    \label{fig:enter-label}
\end{figure}
\pagebreak
\setcounter{section}{6}
\subsection{Section 6: Micrograph of photonic wire bonds}
\setcounter{figure}{0}
\begin{figure}[h!]
    \centering
    \includegraphics[width=0.8\linewidth,keepaspectratio]{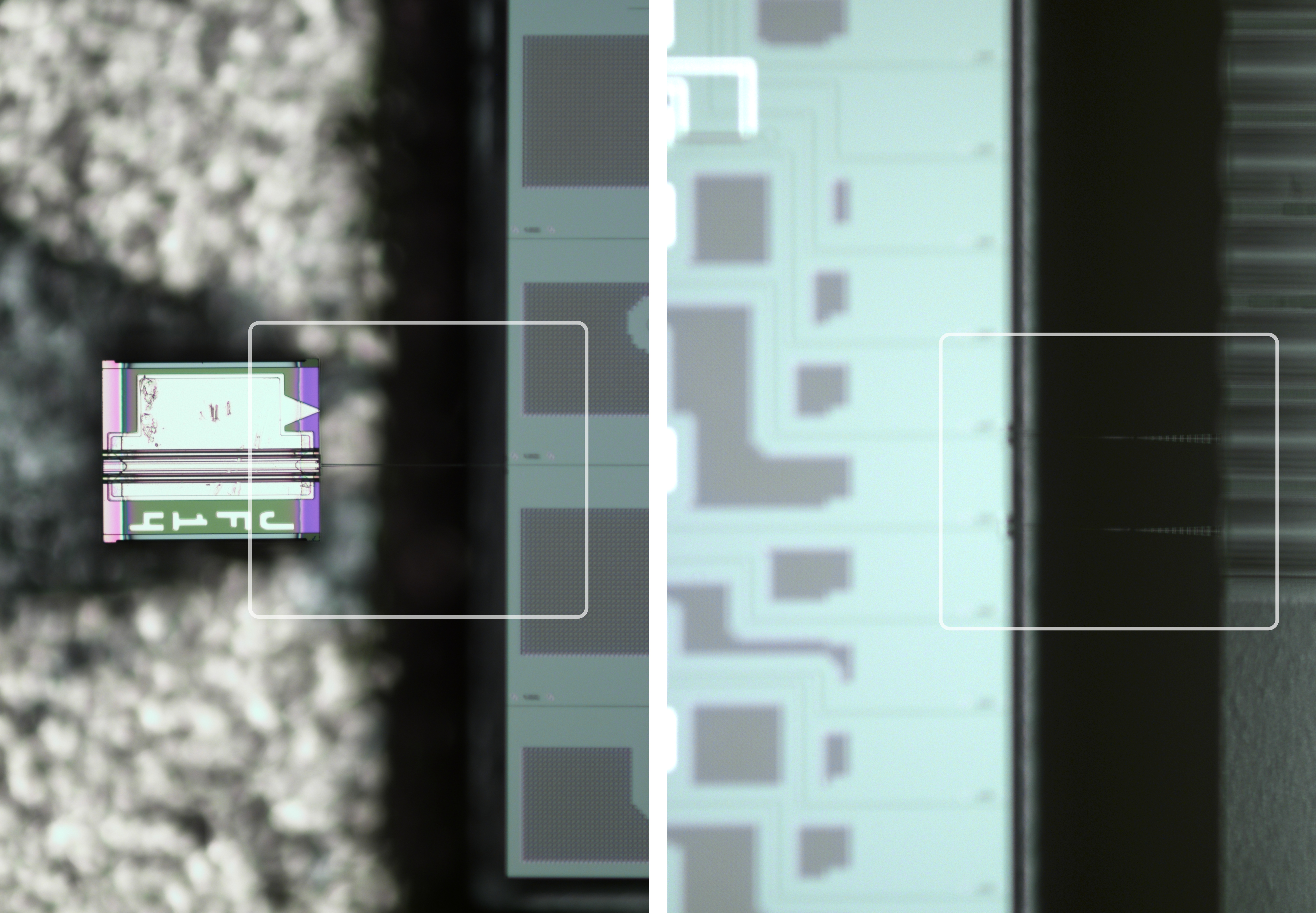}
    \caption{Micrograph of the (Left) laser-to-PIC and (Right) PIC-to-fiber photonic wire bonds.}
    \label{fig:enter-label}
\end{figure}

\end{document}